\documentclass[journal,twoside,web]{ieeecolor}
\usepackage{generic}
\usepackage{cite}
\usepackage{amsthm}

\usepackage{graphicx}          
\usepackage{todonotes}
\usepackage{amsmath,amssymb,amstext}
\usepackage{graphicx}
\usepackage[justification=centering]{caption}
\usepackage{subcaption}
\usepackage{booktabs}
\usepackage{algorithm}
\usepackage{algpseudocode}
\usepackage{color}
\usepackage{textcomp}

\newtheorem*{remark}{Remark}
\newtheorem{theorem}{Theorem}

\newcommand{\R}{\mathbb{R}}

\def\BibTeX{{\rm B\kern-.05em{\sc i\kern-.025em b}\kern-.08em
    T\kern-.1667em\lower.7ex\hbox{E}\kern-.125emX}}
\begin{document}






\title{A Polarization Opinion Model Inspired by 
Bounded Confidence Communications}
\author{Jacek Cyranka and Piotr B. Mucha
\thanks{This paragraph of the first footnote will contain the date on 
which you submitted your paper for review.  The project is financed by the Polish National Agency for Academic Exchange. JC has been partly supported by the NAWA Polish Returns grant PPN/PPO/2018/1/00029.
PBM's work has been partly supported by the NCN Grant No. 2018/30/M/ST1/00340 (HARMONIA).
Both authors have been  supported by IDUB UW grants Nowe Idee.
}
\thanks{Jacek Cyranka is with the Institute of Informatics, University of Warsaw, ul. Banacha 2, 02-097 Warszawa, Poland (e-mail: cyranka@mimuw.edu.pl). }
\thanks{Piotr B. Mucha is with the Institute of Applied Mathematics and Mechanics, University of Warsaw, ul. Banacha 2, 02-097 Warszawa, Poland.}
}

\maketitle

\begin{abstract}                  
We present an opinion model founded upon the principles of the bounded confidence interaction among agents. Our objective is to explain the polarization effects inherent to vector-valued opinions. The evolutionary process adheres to the rule where each agent aspires to increase polarization through communication with a single friend during each discrete time step. The dynamics ensure that agents' ultimate (temporal) configuration will encompass a finite number of outlier states. We introduce deterministic and stochastic models, accompanied by a comprehensive mathematical analysis of their inherent properties. Additionally, we provide compelling illustrative examples and introduce a stochastic solver tailored for scenarios featuring an extensive set of agents. Furthermore, in the context of smaller agent populations, we scrutinize the suitability of neural networks for the rapid inference of limit configurations.
\end{abstract}


\section{Introduction}

Agent-based models have emerged as a valuable tool for understanding nonlinear phenomena in various contexts
\cite{AnInt,Epid,Vote}. By modeling the interactions between agents and their local neighborhoods, these models provide a simple yet effective description of complex dynamics \cite{Vic}. In particular, these models have been used to predict the preferences and opinions of groups of individuals, with applications ranging from social network analysis to political science \cite{Wei,Kou,Yu}. As the model's time horizon increases, the agents are expected to converge to a few distinct states, the phenomenon of partial consensus or polarization of opinions
\cite{Blond,Mucha,Boh}.


This manuscript introduces a model to characterize a collective of agents, each possessing diverse opinions represented as vector-valued entities. These opinions encompass various subjects, encompassing evaluations of moral principles, the desirability of goods, or preferences for fashion trends.
Our primary focus is on scenarios wherein individual agents promote their individualism or radicalism, leading to a polarization of opinions within the entire group. Notably, like in the existing class of models, opinions evolve solely by communication with a single group member during each time step. However, in our model, the selection of this "friend" is driven by the polarization maximization paradigm, resulting in a few radical opinions.

Henceforth, we present a bounded confidence model predicated upon the emergent dynamics of polarization observed within a collective of agents \cite{pol94}, and \cite{alvim2021multi,Naoki22,zha2020opinion}. We draw inspiration from
the Deffuant-Weisbuch \cite{DW00,DW02,Weisbuch04}, and the Hegselmann-Krause \cite{HegKra,Nedic12} systems, along with other relevant works, see \cite{DONG201857} and references therein. By combining these foundational ideas, we aim to present a comprehensive framework for analyzing and understanding polarization dynamics in agent-based systems. As a product, we obtain an attempt to explain the choice of the leading opinions in the groups obeying the polarization maximization paradigm.  Regarding the multi-agent dynamics, the agents' global group reward function (utility function) is defined as the total polarization function. The selection of each friend is motivated solely by the global reward maximization paradigm, studied within machine learning optimization frameworks in the Markov decision process (MDP) setting \cite{sutton,marl-book}. Unlike in the control learning approach, we do not aim to design an optimal agent controller but rather use analytical and numerical methods to directly study the multi-agent dynamical system for reward maximization and its limit. To overcome the curse of dimensionality in finding the optimal friend at each step, we design a stochastic solver shown to converge to the full system with a time-step going to zero.


The model presented herein holds promising potential for explicating the emergence of leaders or leading opinions within groups that actively pursue polarization as a central objective. Within these group dynamics, members are driven by dissimilarity, radicalism, or even a form of individualism. Instances of such behavioral patterns can be observed in various domains, including politics \cite{Balda07,ramirez06}, sport-related communities such as fan clubs, social-networks \cite{bursts}, and even in the realm of fashion \cite{DONG201857,Kashima21}.

The proposed model offers a valuable framework for understanding the underlying mechanisms that contribute to establishing influential figures or dominant viewpoints in these polarized settings. By leveraging the principles of polarization maximization and bounded confidence, the model facilitates the exploration of intricate social interactions and opinion dynamics, shedding light on the complex processes that engender leadership and influential opinions within polarized groups. Consequently, the insights derived from this model may enrich our comprehension of group dynamics across diverse fields and provide a deeper understanding of how polarization shapes the formation of leaders and leading opinions.

With the interpretation removed from the opinion dynamics, we present our findings.
We propose an agent-based model designed to facilitate the self-organization of individuals into distinct groups, with the primary objective of clustering themselves. This clustering process is regulated by a predefined function that operates on the agents' positions, with their positions being fixed in $\mathbb{R}^D$.
Our model can be perceived as a modified iteration of the Deffuant-Weisbuch model \cite{DW00} and the Hegselmann-Krause model \cite{HegKra}, where communication between agents is confined to interactions with only one other agent at each time step, see also \cite{Yang,boundedconf}. Furthermore, the system evolution necessitates an increment in the value of the polarization function, thereby yielding a dynamic process that drives all agents toward a few pivotal points. These points collectively form the vertices of a convex hyper-polygon, which is determined by the initial positions of the individuals.


Our approach is founded on classical mathematical analysis and incorporates modern machine-learning techniques.
We have devised a rapid stochastic solver to handle large clusters effectively and ensure stability. This solver exhibits exceptional scalability, accommodating a considerable number of agents, even up to $10^6$, and completing a single simulation step within seconds.
Additionally, we assess the possibilities of approximating the long-term dynamical limit (inferring the exact attractor based on the agent's initial conditions) by leveraging
the predictive capabilities of neural networks. This task presents substantial challenges, even with a limited number of agents, to be addressed by further research. We evaluate the widely used approach of deep encoder-decoder, where we incorporate two methods of ensuring agents' permutation invariance.

In summary, our work presents a comprehensive and rigorous approach that combines mathematical analysis, machine learning techniques, and efficient computational methods to understand and predict the dynamics of multi-agent systems and their collective polarization behavior.

\section{Model Introduction and Comparisons}
\label{sec:model}

Let us define the models.
Consider an ensemble of $N$ agents. Each agent is specified by its vector-valued position $z_k(t)$ at time $t$, embedded in the $D$-dimensional Euclidean space $\R^D$, and the whole set of positions is denoted by $\bar z(t)$. Hence for fixed  time $t$ we have:
\begin{equation}\label{zz}
\begin{array}{c}
     z_k(t)=(z_k^1(t),z_k^2(t),...,z_k^D(t)) \in \R^D \mbox{ \ \ for  \ \ } k=1,...,N; \\[7pt]
     \mbox{ \ with \ }
    \bar z(t)=\{z_1(t),...,z_N(t)\} \in \R^{D\times N}.
    \end{array}
\end{equation}
We consider the discrete-time dynamics with time step $\Delta t>0$, but at the analysis stage, we will consider the continuous limit for $\Delta t \to 0$.

Alternatively, one can consider our set on a graph $G=(V,E)$, where $V=\{v_1,...,v_N\}$ represents the vertices corresponding to labels of agents   and $E$ edges of graph determine communication between vertices, so $(i,j)\in E$, then 
 $i$-th agent sees the $j$-th. The communication is not symmetric in the considered case, so $(i,j)\in E$ does not imply $(j,i)\in E$. Our graph is directed. At each vertex $v_k$ for $k=1,...,N$ we define the position/opinion $z_k \in \R^D$.
 And if $(i,j) \in E$, then vertex $v_j$ influences vertex $v_i$.
 The dynamics change the communication in time; hence, we write $G(t)=(V,E(t))$.
The structure of the graph naturally 
defines the communication matrix
$C(G) \in M_N^N$:
\begin{equation}
    \{C(G)\}_{ij}=\left\{ 
    \begin{array}{ll}
        1 &  \mbox{ if } (i,j) \in E\\
        0 &  \mbox{ if } (i,j) \notin E
    \end{array}
    \right.
\end{equation}

The language of graphs allows a better understanding of communication between agents.
The edges $E(t)$ show the directed link between vertices, expressing which agent influences others from the group. In our case, the graph description clearly explains the phenomenon of clustering of the whole group. When using the notation of the communication matrix, the clustering means that $C(G)$, after a suitable perturbation of indices,
becomes a block matrix.

The dynamics of the agent's ensemble is studied over time under the following polarization paradigm denoted as $(P)$:

\smallskip 

\emph{The objective is to maximize the sum of distances between all agents over time, subject to interactions that only permit attraction.}

\smallskip 

Here, the term "distances" can be interpreted as a generalized function of distances between the agents' positions/opinions. The basic interpretation of our model is an opinion dynamics maximizing the polarization. Each agent has its own opinions described by the vector-valued state $z_k$.

Having definition (\ref{zz}), we introduce the polarization function agreeing with paradigm (P)
\begin{equation}\label{eq:loss}
    L(\bar z) =\frac{1}{N^2} \sum_{k,l=1}^N g(z_k-z_l).
\end{equation}
Above $g:\R^D \to \R_+ \cup \{0\}$, at least $g \in C^2$. The main examples are $g(w)=|w|$ and $g(w)=|w|^2$ for $w \in \R^D$ with notation $$\displaystyle |w|=\sqrt{\sum_{k=1}^D (w^k)^2}.$$
 The basic assumption for the needs of this note is that $g(\cdot)$ is convex and
 even $g(-w)=g(w)$.

Determining the dynamics requires restricting the kind of interactions between the agents.
Considering a group of agents labeled from $1$ to $N$, each possessing vector-valued opinions denoted as $z_k \in \mathbb{R}^D$. To describe the evolution of opinions, we define the following update rule:
\begin{equation}\label{CM}
z_k(t+\Delta t)= z_k(t) + \Delta t \big(z_{l(k)}(t) - z_k(t)\big).
\end{equation}
Here, at each time step, each agent changes its opinion by considering only the opinion of one agent in the group, labeled as $l(k)$. This "friend" is selected to maximize a polarization function $L$ for the entire group.
In general, an agent can communicate only with itself.

A fundamental requirement is that $L$ can be expressed as a sum of individual factors $L^k$ for each agent, leading to the following expression:
\begin{equation}\label{L-ass}
L(\bar z) = \frac{1}{N}\sum_{k=1}^N L^k(\bar z)=\frac{1}{N}\sum_{k=1}^N \left(\frac{1}{N}\sum_{l=1}^N g(z_k-z_l)\right),
\end{equation}
where $g$ is a chosen increasing function. The local choice of maximization for the $k$th agent aligns with the overall increase in the full polarization function $L$.

We provide an interpretation for the "friend" $l(k)$: it can be perceived as the current leader for the $k$th member of the group. The $k$th agent seeks to adopt the opinion of the $l(k)$th agent to maximize the $L^k$ component of the polarization function. It is crucial to note that the identity of the leader $l(k)$ is subject to change over time.


%

In the real case, it is difficult to assume that each agent has the full information about the opinions in the whole group.
To maintain the interpretation of the model in close alignment with the bounded confidence opinion dynamics, we propose the following stochastic modification of (\ref{CM})-(\ref{L-ass}):
\begin{equation}\label{SCM}
z_k(t+\Delta t)= z_k(t) + \Delta t \big(z_{l_S(k)}(t) - z_k(t)\big),
\end{equation}
Here, the selection of $l_S(k)$ is determined by a random sample $A_S=\{m_1,...,m_S\}$ drawn from the set of all indices $\{1,...,N\}$. The purpose of this stochastic approach is to maximize the stochastic approximation of the polarization function for the $k$th agent, given by:
\begin{equation}\label{SL-ass}
L^k_S(\bar z) = \frac{1}{S}\sum_{l\in A_S} g(z_k-z_l).
\end{equation}
In this case, the agent considers the opinions of a random subset of neighbors and chooses the local leader to maximize  function (\ref{SL-ass}).

This stochastic version appears to be more realistic than browsing the opinions of the entire group. Introducing randomness in the selection process reflects the inherent uncertainty and limited information that individuals often have in real-world opinion dynamics. A smaller, random sample of neighbors influences the agent's decision-making, mirroring how individuals in social networks often interact with only a subset of their connections rather than the entire network. This modification better captures the dynamics of opinion formation and polarization in realistic social scenarios.

\bigskip 

As our model is introduced, we can begin the comparison of our approach with known ones. We shall start with the employed definition of polarization central to defining the agents' global dynamics. We refer here to papers \cite{pol94} and \cite{Naoki22}, where similar way of measurement of polarization has been introduced. In \cite{Naoki22} we find a simplified version from \cite{pol94} in the following form
\begin{equation}
    L(\bar z) \sim \sum_{i,j=1}^N a_{ij}|z_i-z_k|,
\end{equation}
where $a_{ij}$ are some weights corresponding to the sizes of neighborhoods of the $i$th and $j$th agent. Thus, our definition (\ref{eq:loss}) follows that rule.


\smallskip 

In terms of the model dynamics, our model can be compared with the classical Deffuant-Weisbuch (DW) model \cite{DW00}. In the DW model, a random pair ${k,j}$ is chosen, and their interaction is defined as follows:
\begin{equation}\label{DW}
\begin{aligned}
z_k(t+\Delta t) &= z_k(t) + \Delta t(z_j(t) - z_k(t)), \\
z_j(t+\Delta t) &= z_j(t) + \Delta t(z_k(t) - z_j(t)).
\end{aligned}
\end{equation}
This interaction takes place provided that the opinions $z_k(t)$ and $z_j(t)$ are not too far from each other, i.e., $|z_k(t)-z_l(t)| \leq \sigma$, where $\sigma$ represents a certain interaction radius. In the literature \cite{ZHANG13}, one can find an asymmetric generalization of the DW model.
Comparing to (\ref{CM}), the crucial distinction lies in the choice of the friend for agent $k$, denoted as the $l(k)$th agent. In our model, this choice is determined by the polarization function. The friend selection is guided by $L$ maximization. Although, at any given moment $t$, the friend $l(k)$ may be far away from agent $k$, it is envisioned as the final position that the agent aims to reach in the long run. Consequently, for large time intervals, the distance between the opinions of the $k$th and $l(k)$th agents is expected to be small. This reflects the tendency of agents to converge toward specific opinions over time as dictated by the polarization function, even if their initial opinions are distant from each other.


The second natural comparison is with the Hegselmann-Krause (HK) model \cite{HegKra}. Let us recall the dynamics for HK in the following setting:
\begin{equation}\label{HK}
z_k(t+\Delta t)=z_k(t) + \frac{\Delta t}{\sharp \,\mathcal{N}(k,t)} \sum_{j\in \mathcal{N}(k,t)} (z_j(t)-z_k(t)),
\end{equation}
In this model, the interaction takes place among members of the subgroup 
$$\mathcal{N}(k,t):=\{ l : |z_k(t)-z_l(t)| \leq \sigma\},
$$
where $\sigma$ represents the interaction radius, and $\sharp \,\mathcal{N}(k,t)$ denotes the number of elements in the set $\mathcal{N}(k,t)$. Each agent aims to reach a state that is an average of its neighbors, limited by the distance $\sigma$. Importantly, communication with more than one agent is allowed in this model, making it asymmetric.
The interpretation of this model is as follows: the set $\mathcal{N}(k,t)$ defines the expected state of the agent $k$. It can be seen as a multi-agent generalization of the single case of the friend $l(k)$ in our proposed model (\ref{CM})--(\ref{L-ass}). This connection with the HK model suggests an interesting direction for possible extensions of the class of models (\ref{CM})--(\ref{L-ass}).

Finally, let us mention the system from \cite{Manifold}. 
There, the model aims at polarization of opinions, where dynamics is given as the gradient flow
\begin{equation}\label{grad-flow}
    \dot z_k \sim - \partial_{k} L(\bar z),
\end{equation}
albeit with a modification, i.e., the whole model is defined over a manifold. The communication between agents driven by (\ref{grad-flow}) is more involved than our scheme (\ref{CM}).
Also, the very polarization function $L$ in \cite{Manifold} is of a different structure than ours $L$ (\ref{eq:loss}) due to different aims of both approaches. Note that (\ref{grad-flow})
does not correspond directly to the dynamic (\ref{CM}), although they follow the same paradigm, i.e., maximize the polarization function $L$.

The key difference between our model (\ref{CM}) and the models (\ref{DW}) and (\ref{HK}) lies in the fact that our approach inherently leads to the polarization of opinions. In contrast, in the cases of DW and HK models, it depends on the data and parameters of the models \cite{Cons13,Lor10}.
In all cases, we did not restrict the number of possible final clusters of opinions. However, our model's final configuration consists exclusively of the most radical opinions (found on the convex hull of the agents' configuration). This outcome is a consequence of the simple form of our model and the specific choice  of the function $L$ given by (\ref{L-ass}).

\section{The Dynamics}

Determining the rule of motion for an ensemble is a fundamental task in studying dynamics. We require the change of $z_k(t)$ in time to be limited to a convex combination with other agents. Letting the time step $\Delta t \to 0$ in the discrete update rule (\ref{CM}), we obtain an ordinary differential equations (ODE) system.
Hence, we obtain the agents' continuous dynamical system
\begin{equation}\label{ev-rule}
    \dot z_k=-(z_k -z_{l(k)}).
\end{equation}
Analysis of (\ref{ev-rule}) in some cases is simpler, and the features of the rule (\ref{CM}) give 
the same result for small $\Delta t$. The consequence is that $l(k)$ is piece-wise constant in time.

Formally, we solve (\ref{ev-rule}) for all possible choice of $l(k)$, and then by maximizing polarization function $L$, 
we obtain the proper value of $l(k)$.
Hence  the analytical optimization procedure will define the form of communication matrix $C(G(t))$, avoiding the necessity of studying $G(t)$ directly.

Now, we aim to find a proper approach to our models.
The key problem is determining function $l(k)[t]$. Note that the configuration of agents is changing in time, hence the choice of $l(k)$ is also variable in time, as denoted by $[t]$.
Applying the greedy selection of the friend per agent leads to the curse of dimensionality problem. For instance, if we consider a system with ten agents, there would be $10^{10}$ potential interactions to analyze at each time step. There are three potential solutions to address this challenge:

\begin{enumerate}
\item Firstly, we could compute all the $ L $ derivatives analytically to obtain an explicit formula for $l(k)[t]$. 

\item Alternatively, we could employ a machine learning technique using a training set computed through suitable simulations within a machine learning framework. We could then discover the rule of the dynamic and identify the formula for $l(k)$ using this method. 

\item Lastly, we could employ a hybrid approach, where we perform mathematical analysis to the extent possible and use stochastic methods to determine quantities that cannot be computed in a deterministic way.

\end{enumerate}

The last approach seems to be the most efficient in the general case in the form of (\ref{eq:loss}).
Then under the rule of evolution (\ref{ev-rule})
we want to determine $l(k)$. Assuming that $L(\bar z(t))$
is regular enough, say $g \in C^2$, we compute the derivative with respect to time
\begin{multline}\label{grad}
    \frac{d}{dt} L(\bar z(t))=\frac{1}{N^2} \sum_{k,m=1}^N \frac{d}{dt}
    g(z_k(t)-z_m(t))\\
    =\frac{2}{N^2}\sum_{k,m=1}^N \partial_z g(z_k-z_m)\cdot (z_{k(l)}-z_k)\\
    =\frac{2}{N^2}\sum_{i=1}^D\sum_{k,m=1}^N \partial_{z^i} g(z_k-z_m)\cdot (z_{k(l)}^i-z_k^i),
\end{multline}
where $\partial_z g=(\partial_{z^1}g,..., \partial_{z^D}g)$ is the gradient of $g$.
For the above formula we easy deduce that in the discrete case 
\begin{multline}
     L(\bar z(t+\Delta t))-  L(\bar z(t))=\\
    \frac{2}{N^2}\sum_{k,m=1}^N \partial_z g(z_k-z_m)\cdot (z_{k(l)}-z_k) \, \Delta t \\ + O(\Delta t)^2 \mbox{ \ \ for a sufficiently small } \Delta t.
\end{multline}
Hence in order to maximize $L$ we look at  the gradient (\ref{grad}).
This leads to the explicit formula for $l(k)$
\begin{equation}\label{def-lk}
    l(k)={\rm arg^*max}_j \left( \frac{1}{N}\sum_{m=1}^N \partial_z g(z_k-z_m)\right) \cdot (z_j-z_k).
\end{equation}
In the generic case ${\rm arg^*max}_j =
{\rm arg\;max}_j$ as the right-hand side is uniquely determined. Otherwise, ${\rm arg^*max}_j$
selects the smallest index among all minimizers. Note that from that reason the system is very unstable,  since a small perturbation can change significantly the choice of the communication for the whole set.

The structure of the above formula is simple, the vector 
\begin{equation}\label{theta}
    \theta[k,t] = \frac{1}{N} \sum_{m=1}^N \partial_z g(z_k(t)-z_m(t))
\end{equation}
is multiplied by $z_j-z_k$. We look for the maximum of this product over the set $z_j$. The maximizer can be found only at the verities of the convex hull of the set $\bar z=\{z_1,...,z_N\}$. We define:

\begin{itemize}
\item $CH(\bar z)$: the set of all points being a convex combination of points in $\bar z$ and  

\item $VCH(\bar z)$: the set of vertices at the boundary of $CH(\bar z)$.
\end{itemize}

Finally, we return to the graph language, and we say that 
$$
(i,j) \in E(t) \mbox{ \ \ iff \ \ } j = l(i)[t].
$$
It means that only one edge can start at each vertex.

In a special case for $g(w)=|w|^2$, the formula (\ref{def-lk}) takes the simple form
\begin{equation}
    l(k)={\rm arg^*max}_j \big\{(z_k-\{z\})\cdot (z_j-z_k)\big\},
\end{equation}
where 
$\{z\}$ defines the mass center of $\bar z$:
\begin{equation}
    \{z\}(t) = \frac{1}{N} \sum_{k=1}^N z_k(t).
\end{equation}
The formula in question possesses an elegant and insightful geometric interpretation for the planar case $(D=2)$. Precisely, by examining the line spanned by the vector $z_k-\{z\}$, we can ascertain the solution at an element of the boundary of the convex hull and intersection with this line. This example explains the possible lack of the unique choice of ${\rm arg\;max}_j$. It happens when the mentioned line crosses the element of the convex hull precisely in the middle, we can not exclude such a situation. 

\smallskip

{\bf Deterministic dynamics.}
Having the formula for $l(k)$ the dynamical system of our model 
$$
\bar z(t) \mapsto \bar z(t+\Delta t)
$$
can be defined finally. We apply the following procedure

\begin{itemize}

\item -- let $l(k)[t]$ be defined for the $z_k(t)$ at time $t$;

\smallskip 

\item -- let $\displaystyle\theta[k,t]=\frac{1}{N}\sum_{m=1}^N \partial_z g(z_k(t)-z_m(t))$;

\smallskip 

\item -- then $l(k)[t]=
\underset{j\in VCH(\bar z(t))}{\rm arg^*max} \theta[k,t] \cdot (z_j(t)-z_k(t))$;

\smallskip 

\item -- then $z_k(t+\Delta t)=(1-\Delta t) z_k(t) + \Delta t \,z_{l(k)}(t)$.

\end{itemize}
The dynamics of the system have been defined.
For $g$ that is convex, we can demonstrate the following result, which characterizes the structure of the attractor, at least in this case.

\smallskip

\begin{theorem}
\label{thm:convex}
Let $g$ be a convex even function. Let $z_k(0) \in \R^D$ for $k=1,...,N$. Let $\Delta t$ be such that $0<\Delta t <1$. Then $L(\bar z(t))$ tends to a local maximum as $t \to \infty$.
\newline
Moreover  there exist
$N_\infty$ disjoint sets of indices $I_1, ..., I_{N_\infty}$ summing to the whole set $\{1,...,N\}$ such that
$$
\mbox{if $l \in I_k$, then } z_l(t) \to V_k \mbox{ as } t\to \infty,
$$
where $V_1,...,V_{N_\infty}$ is a set of vertices on the boundary of the convex hull $CH(\bar z(\infty))$.

{\rm [Clustering]} Finally, there exists time $t^*>0$ such that 
$$
\mbox{if \ } (k,m) \in E(t^*), \mbox{ \ then \ }
(k,m) \in E(t) \mbox{ \ and \ }
k,m \in I_p
$$
for some $p \in \{1,..., N_\infty\}$ and all $t>t^*$.

%
\end{theorem}
{\bf Proof.} 
%
Note that the function $L(\bar z(t))$ is an increasing function on the solution. Furthermore, $L(\cdot)$  is bounded, since the rule (\ref{CM}) can not increase the diameter of the whole set $\bar z$. Therefore, $L(z(t))$ serves as the Lyapunov function for our system, indicating that the solutions tend to a stationary configuration where $l(k)=k$ for each $k$. Analysis of (\ref{def-lk}) reveals that the stabilization holds if $l(k)$ is an index of the element from the convex hull, or $\theta[k,t]=0$. This condition holds since we are at the minimum of the function, utilizing the convexity of $g(\cdot)$. Therefore, the choice of $l(k)$ from the convex hull increases $L$ when we are not at the equilibrium. 
Since the polarization function $L$ is bounded, $L(\bar z(t))$ reaches its local maximum for $t\to \infty$.

In order to show the structure of the attractor we need to use essentially convexity of the $L$
with respect to $z_k$, taken as coordinates. 
We need to demonstrate that the final shape of the set is determined by a few agents from the initial configuration. Note that the dynamics guarantee $$CH(\bar z(t+\Delta)) \subset CH(\bar z(t)),$$
ensuring that $CH(\bar z(\infty))$ is a well-defined non-empty convex set. 

Since $L(\bar z(t))$ tends to a local maximum, then by definition we deduce (\ref{def-lk}) 
$$|z_k(t)-z_{l(k)}(t)| \to 0
\mbox{ \ as \ } t\to \infty.
$$
Since from a suitable moment, the time step $\Delta t$ is significantly smaller then 
the convergence rate, it follows that agents need to group to a few clusters.
It means that all $\{z_k(t)\}_{t}$ converge to some points and we name them $V_k$ for $k=1,...,N_\infty$. 
By definition, these are vertices of the boundary of $CH(\bar z(\infty))$. 

The part concerning the clustering follows directly from
the first part concerning the limit points $V_1,...,V_\infty$. Disjoint sets of indices $I_1,...,I_{N_\infty}$
define the perturbation of indices of the agents in such a way that the communication matrix $C(G(t))$
takes the following block form
\begin{equation}\label{C-Gt}
    C(G(t))=\left(
    \begin{array}{cccc}
        \Pi_1 & 0 & \dots & \\
        0 &  \Pi_2 & 0 & \dots \\
         & & \ddots & \\
         & \dots & 0 & \Pi_{N_\infty}  
    \end{array}
    \right),
\end{equation}
where $\Pi_k$ is a square matrix
such that in each row has only one nonzero element.
The clustering essentially means that the form (\ref{C-Gt}) of $C(G(t))$ 
does not change for $t> t^*$. 
$\Box$



\section{Stochastic Model}
\label{sec:stochas}
When dealing with a large number of particles, the formula given by (\ref{theta}) may not be suitable as it requires considering all particles. However, it is essential to note that we must still choose one agent with which the considered particle interacts. Through analysis, we know that such an agent is located necessarily at the boundary of the convex hull. We substitute the general law with an approximation which will address the central paradigm as a solution.

The central issue is to efficiently calculate $l(k)$. As we observe from (\ref{def-lk}), the the quantity $\theta[k,t]$ from (\ref{theta}) plays a vital role in the computation process $l(k)$.

According to the definition (\ref{def-lk}), $l(k)$ can only take natural values. As a result, it is piece-wise constant with respect to time, and due to the structure of the formula, the number of possible values is limited to several dozen values, even for systems with several million agents. Consequently, we can compute or approximate $l(k)$'s through random approximation methods.

Given $S \in \mathbb{N}$ and $S \leq N$ we define
\begin{equation}
\label{eq:stocheq}
    \theta_S(z_k)=\frac{1}{S} \sum_{m\in A_S} \partial_z g(z_k-z_m),
\end{equation}
where $A_S=\{m_1,...,m_S\}$ is a random sub-sample from the set $\{1,...,N\}$.
The Central Limit Theorem entails
\begin{equation}
    P(|\theta(z_k)-\theta_S(z_k)|< \epsilon) \sim \frac{C(\epsilon)}{\sqrt{S}}.
\end{equation}
When $S$ is relatively small, the convergence rate is less than satisfactory. Nonetheless, our objective is to demonstrate that $\theta_S \to \theta$, not through increasing $S$.

{\bf Stochastic dynamics.}
Then we redefine our dynamics as follows:
\begin{itemize}

\item -- let $l_S(k)[t]$ be defined for the $z_k(t)$ at time $t$;

\item -- given $S$, we take a random sample $A_S$ of indices;

\item -- let $ \displaystyle \theta_S[k,t]=\frac{1}{S}\sum_{m\in A_S}\partial_z g(z_k(t)-z_m(t))$;

\item -- then $l_S(k)[t]=\underset{j \in VCH(t)}{\rm arg^*max} \theta_S[k,t] \cdot (z_j(t)-z_k(t))$;

\item -- then $z_k(t+\Delta t)=(1-\Delta t) z_k(t) + \Delta t \,z_{l_S(k)}(t)$.

\end{itemize}

We  prove the following theorem:

\smallskip 

\begin{theorem}
Given $N$ and $S \leq N$, suppose that the selection of $l(k)[t_0]$ remains stable under small perturbations of $\bar z(t_0)$. In such a scenario, there exists $t_1 > t_0$ such that the following holds:
%
\begin{equation*}
    P(\big\{l(k)[t]=l_S(k)[t]\big\}) \to 1 \mbox{ \ \ as \ \ } \Delta t \to 0
\end{equation*}
for all $t \in (t_0,t_1)$.
\end{theorem}

{\bf Proof.}
To establish the aforementioned result, we must make several observations. Assuming that $g \in C^2$ and the second derivative is bounded, we can fix $\Delta t_0$ to be sufficiently small. This leads us to:
%
\begin{multline}
    |\partial_z g(z_k(t+\Delta t)-z_m(t+\Delta t))-
    \partial_z g(z_k(t)-z_m(t))| 
    \\
    \leq C\Delta t
\end{multline}
for $\Delta t \leq \Delta t_0$. And consequently 
\begin{multline}
    |\partial_z g(z_k(t+p\Delta t)-z_m(t+p\Delta t))-
    \partial_z g(z_k(t)-z_m(t))| 
    \\ \leq Cp\Delta t \mbox{ \ \ for  \ \ }
    p\Delta t \leq \Delta t_0.
\end{multline}
The proof will be based on the idea of splitting the time step. The dynamics of the set implies that $l(k)[t]$ is locally constant for $t \in [t_0, t_1)$, where $t_1 > t_0$. This behavior is guaranteed by the assumption of stability at $t_0$, i.e., the choice of $l(k)$ by formula (\ref{def-lk}) is stable under small perturbations of the data. As a result, $l(k)$ is uniquely determined, and there exists a small $\delta > 0$ such that all $Z \in B(0, \delta) \subset \mathbb{R}^D$ it holds that  
\begin{multline}\label{ll2}
    l(k)[t]={\rm arg^*max}_j \theta[k,t] \cdot (z_j(t)-z_k(t)) 
    \\ = 
    {\rm arg^*max}_j (\theta[k,t]+Z) \cdot (z_j(t)-z_k(t)).
\end{multline}
To maintain the value of $l(k)$, it is necessary to ensure that the precision of the computation of $\theta$ is accurate up to a correction term of magnitude $\delta$.
Note that for $p\Delta t \leq \Delta t_0$
\begin{equation}
    |\theta[k,t+l\Delta t]-\theta[k,t]| \leq C|\Delta t_0| \mbox{ \ for \ } l=1,...,p.
\end{equation}
Then as $C|\Delta t_0|\leq \delta/2$ and
\begin{multline}\label{theta-stab}
    l(k)[t+l\Delta t]={\rm arg^*max}_j \, \theta[k,t] \cdot (z_j(t)-z_k(t)) \\ \mbox{  \ for \ }
    l=0,...,p.
\end{multline}
Hence for $p$ steps with times $t+\Delta t, ..., t+ p\Delta t$ with $p\Delta t \leq \Delta t_0$, quantity $\theta$ is computed just one time. 

Having (\ref{theta-stab}), we take  the stochastic model. The central point then is the computation, or rather stochastic approximation of $\theta[k,t]$. Since at each time step we take 
a sample of order $S$ and to fill the whole $\Delta t_0$ we make $p$ steps. 

The Central Limit Theorem says that
\begin{equation}\label{prob}
    P(|\theta_{pS}-\theta|< \delta/2) \geq 1-C(\sqrt{pS})^{-1}.
\end{equation}
for a sample of order $pS$.
Hence it follows by (\ref{ll2}) that with probability at least $1-C(\sqrt{pS})^{-1}$
\begin{multline}\label{ll5}
   l(k)[t+l\Delta t]={\rm arg^*max}_j \,\theta_{pS}[k,t] \cdot (z_j(t)-z_k(t)) \\ \mbox{  \ for \ }
    l=0,...,k.
\end{multline}
On the other hand it is clear that the growth of the polarization function in the stochastic model 
is given by the left-hand side of the below inequality:
\begin{multline}
   {\rm max}_j \,\theta_{S_1}[k,t] \cdot (z_j(t)-z_k(t)) + ... 
   \\
   +{\rm max}_j \,\theta_{S_p}[k,t] \cdot (z_j(t)-z_k(t))
   \\
   \geq p \, {\rm max}_j \,\theta_{S_1+...+S_p}[k,t] \cdot (z_j(t)-z_k(t)),
\end{multline}
where $S_1,...,S_p$ denotes the samples for time steps from $1$ to $p$ and $S_1+...+S_p$ is the sum of the samples. Next, we note that the right-hand side corresponds to (\ref{ll5}), hence by 
(\ref{prob}) and (\ref{ll2}), it delivers 
the deterministic $l(k)$.
So as $p\to \infty$, by (\ref{ll2}), we get that convergence is almost sure, since the choice of $\Delta t_0$
is arbitrary. $\Box$



\section{Numerical Simulations}
In order to complement the theoretical content of the paper, we presented an experimental study of the model introduced in Sec.~\ref{sec:model} by employing stochastic simulations and neural networks for predicting the agent's limit set. In this section, we fix the opinions vector dimension $D=2$ for illustrative purposes.

\begin{figure*}[h!]
\captionsetup[subfigure]{justification=centering}
\begin{subfigure}[t]{0.32\textwidth}
\caption{The initial condition and the attractor.}
\vspace{-7pt}
\includegraphics[width=\linewidth]{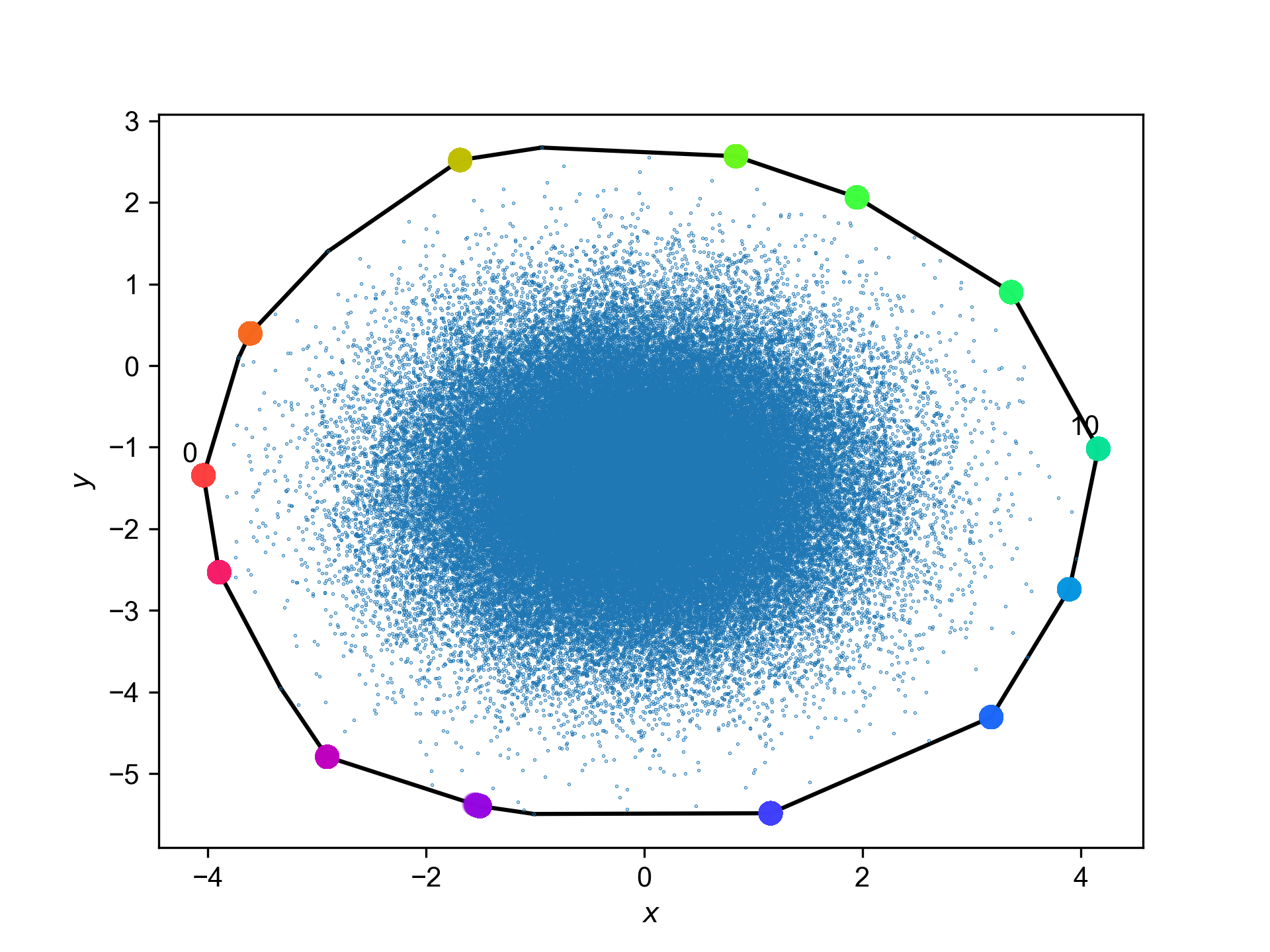}
\end{subfigure}
\begin{subfigure}[t]{0.32\textwidth}
\caption{The loss values w.r.t. the epoch.}
\vspace{-7pt}
\includegraphics[width=\linewidth]{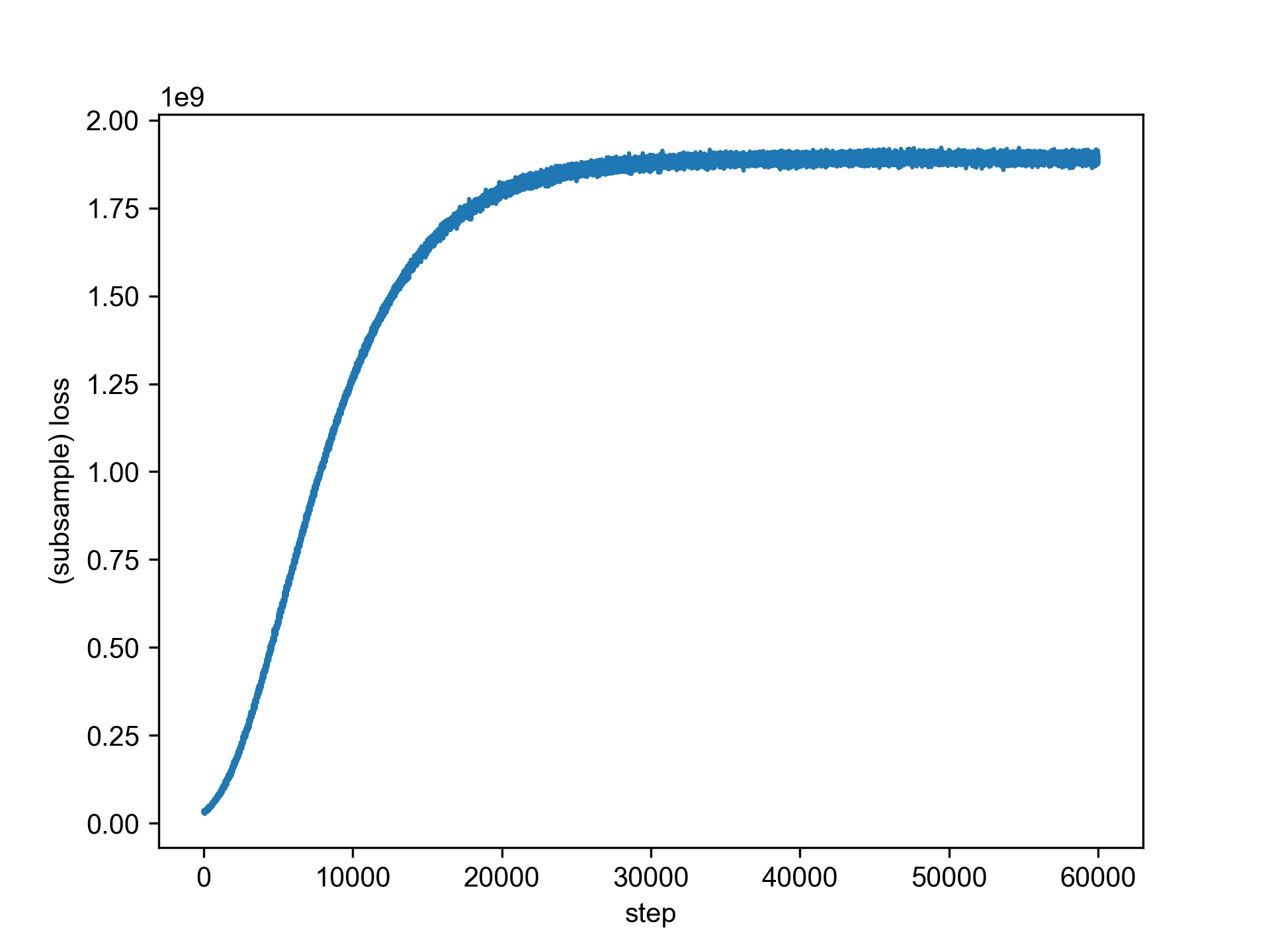}
\end{subfigure}
\begin{subfigure}[t]{0.32\textwidth}
\caption{The attractor density.}
\vspace{-7pt}
\includegraphics[width=\linewidth]{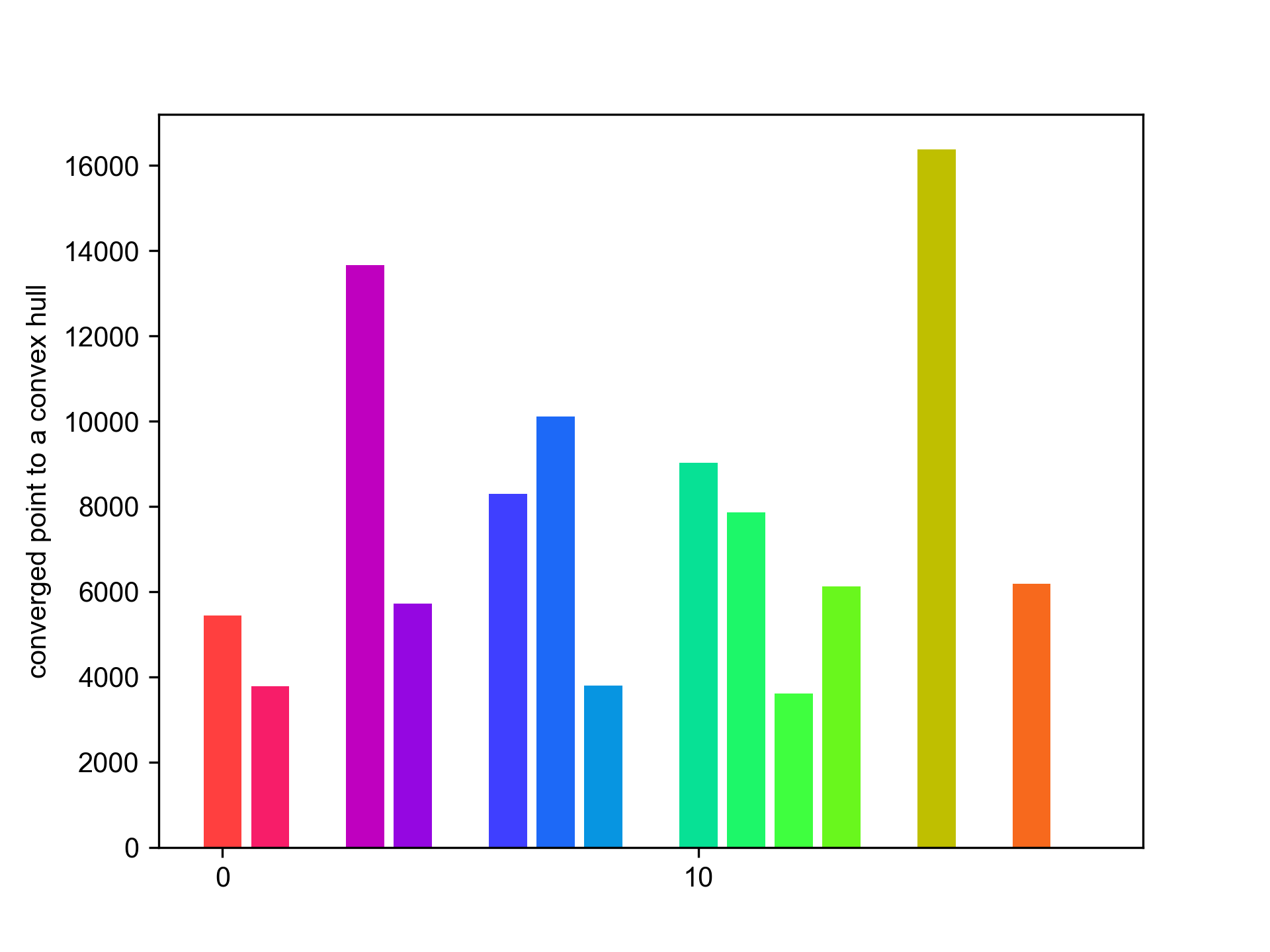}
\end{subfigure}\vspace{-10pt}\\\vspace{-10pt}
\begin{subfigure}[t]{0.32\textwidth}
\includegraphics[width=\linewidth]{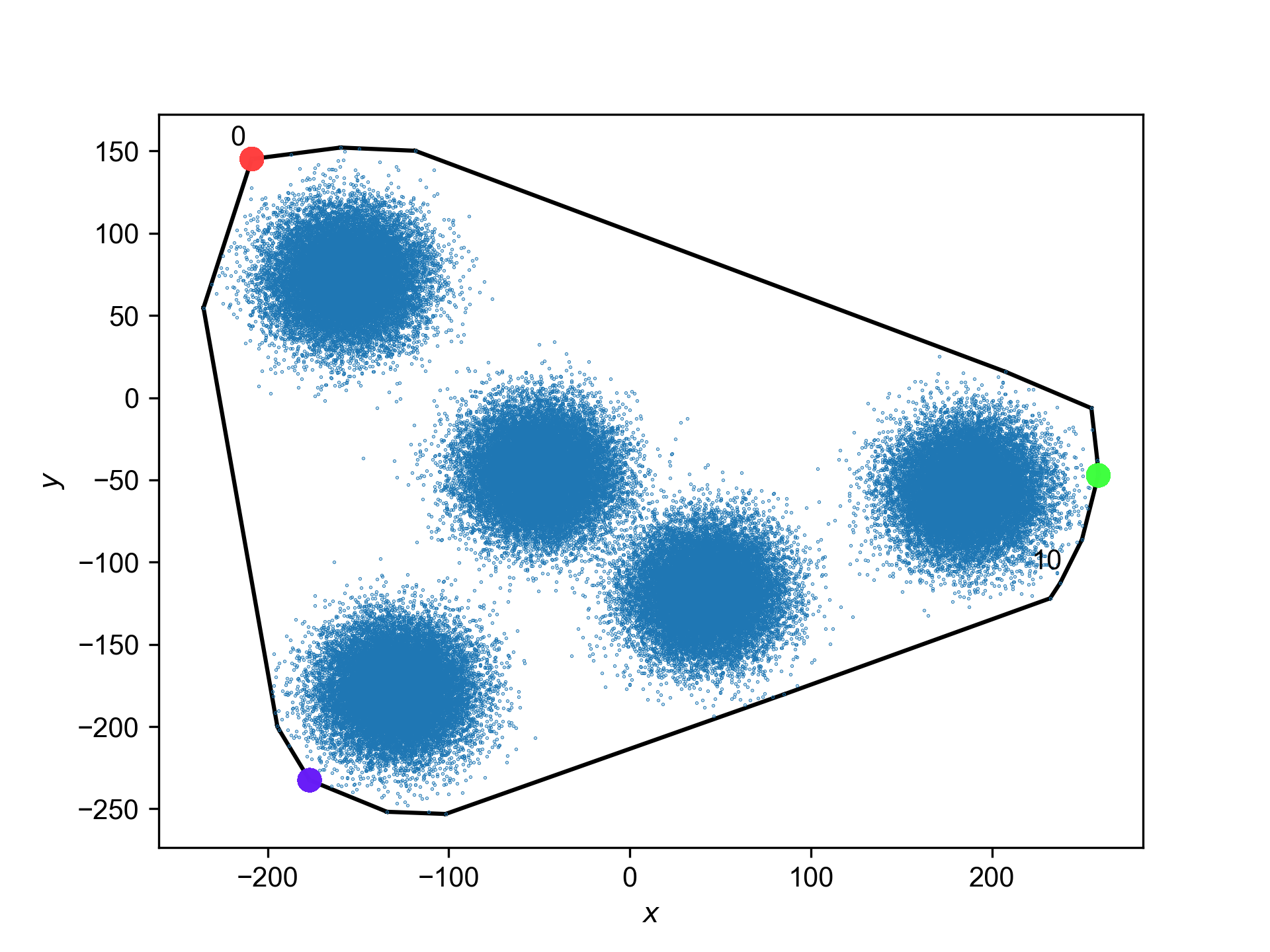}
\end{subfigure}
\begin{subfigure}[t]{0.32\textwidth}
\includegraphics[width=\linewidth]{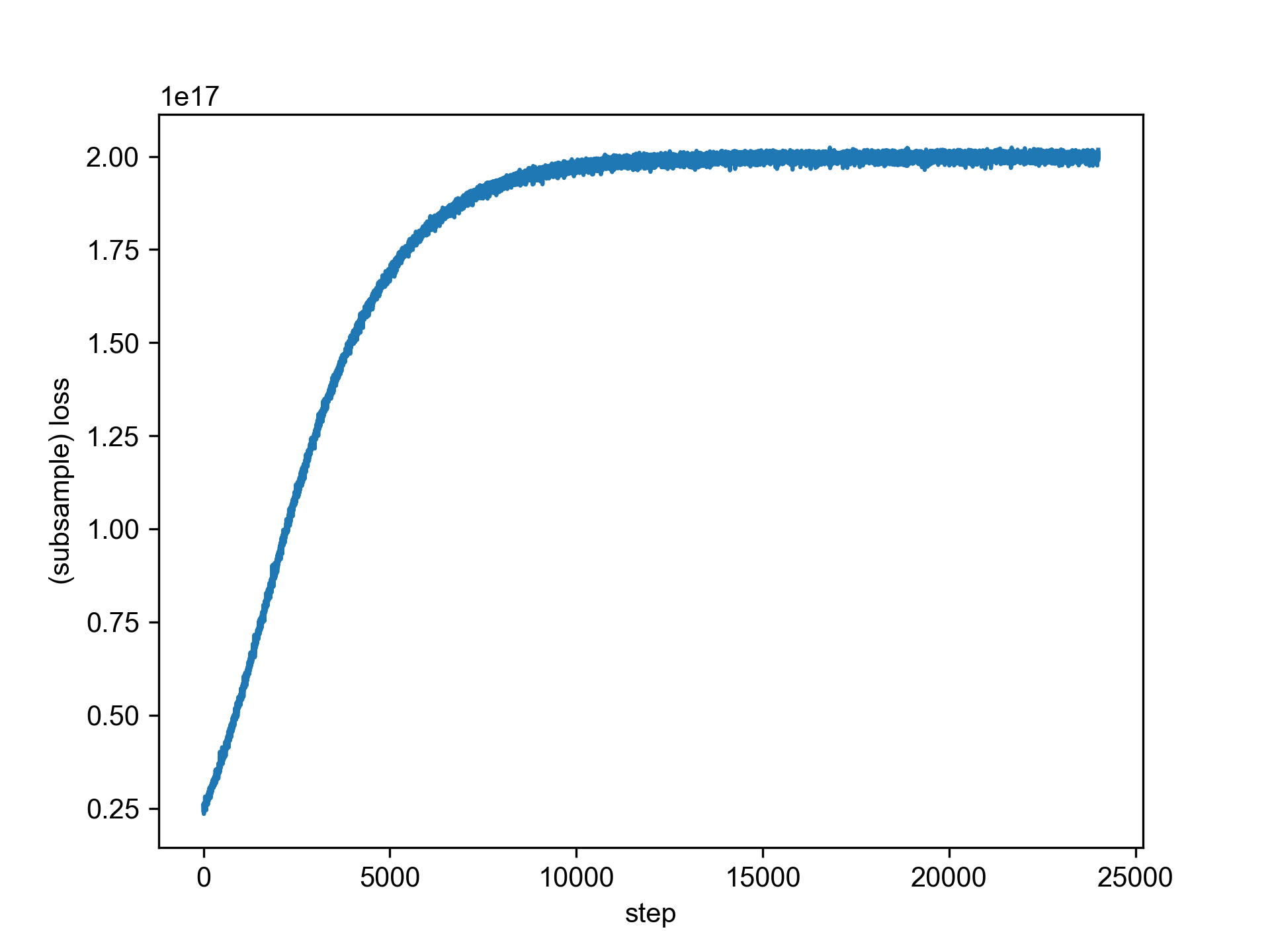}
\end{subfigure}
\begin{subfigure}[t]{0.32\textwidth}
\includegraphics[width=\linewidth]{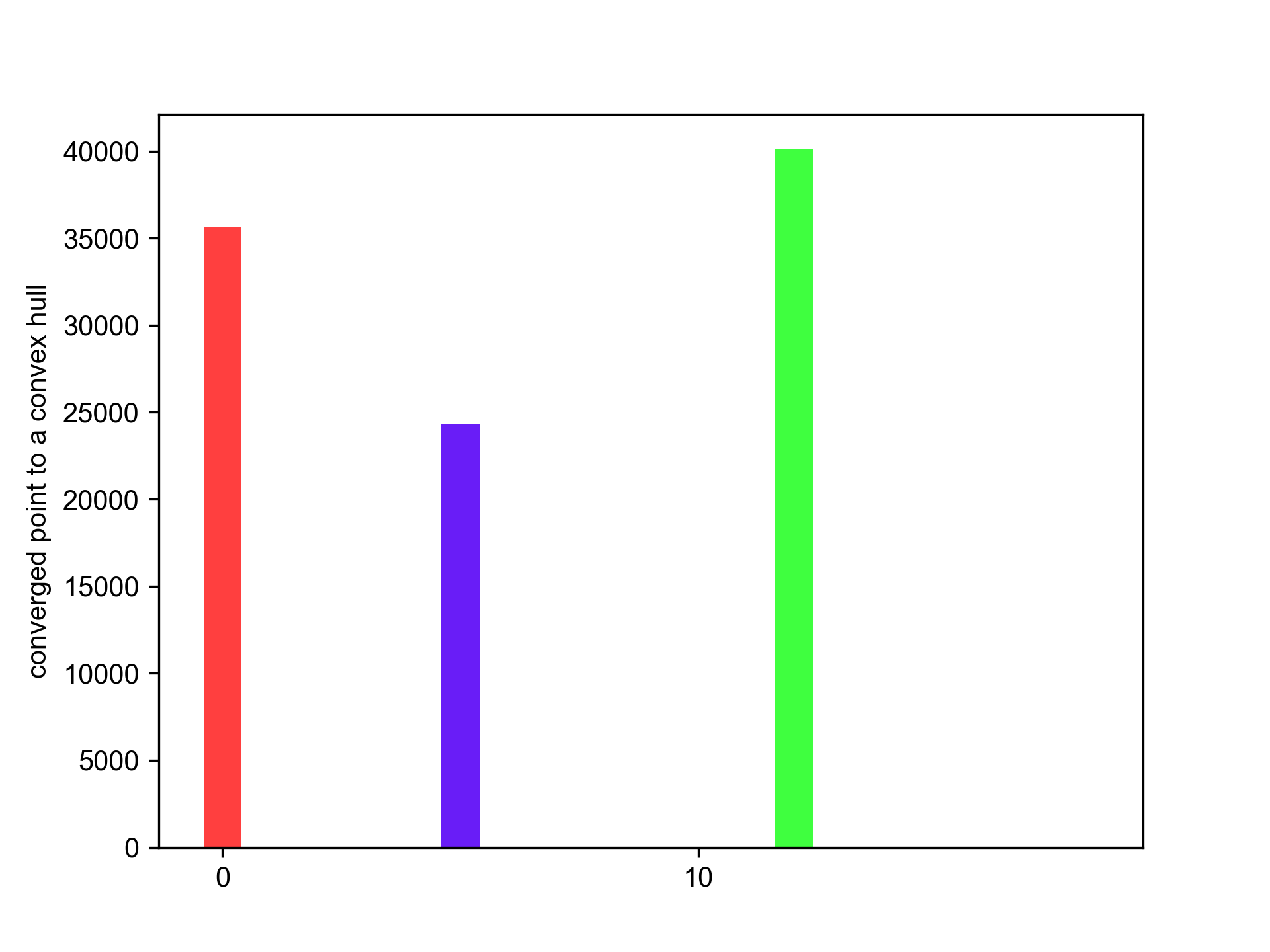}
\end{subfigure}\\
\begin{subfigure}[t]{0.32\textwidth}
\includegraphics[width=\linewidth]{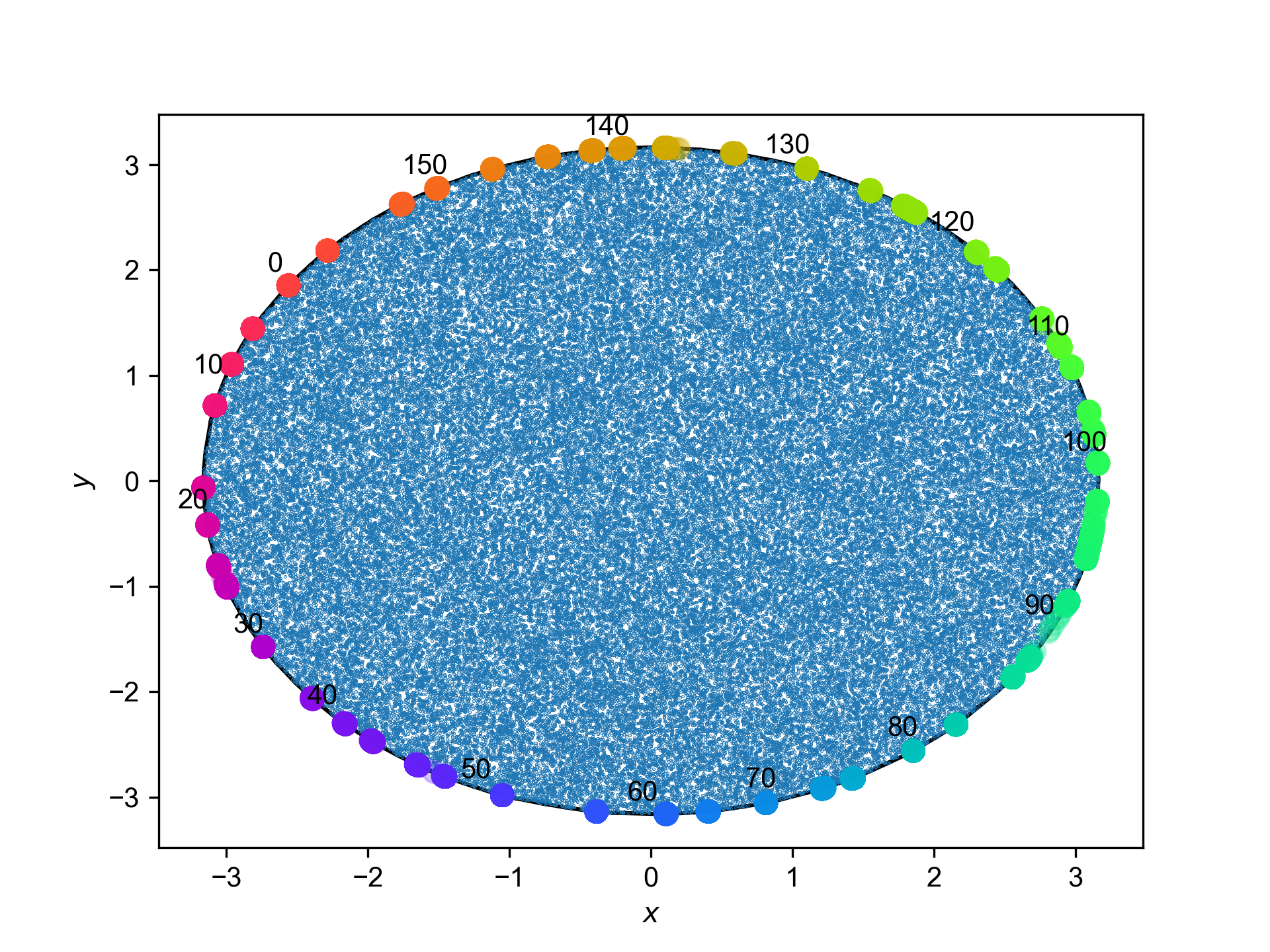}
\end{subfigure}
\begin{subfigure}[t]{0.32\textwidth}
\includegraphics[width=\linewidth]{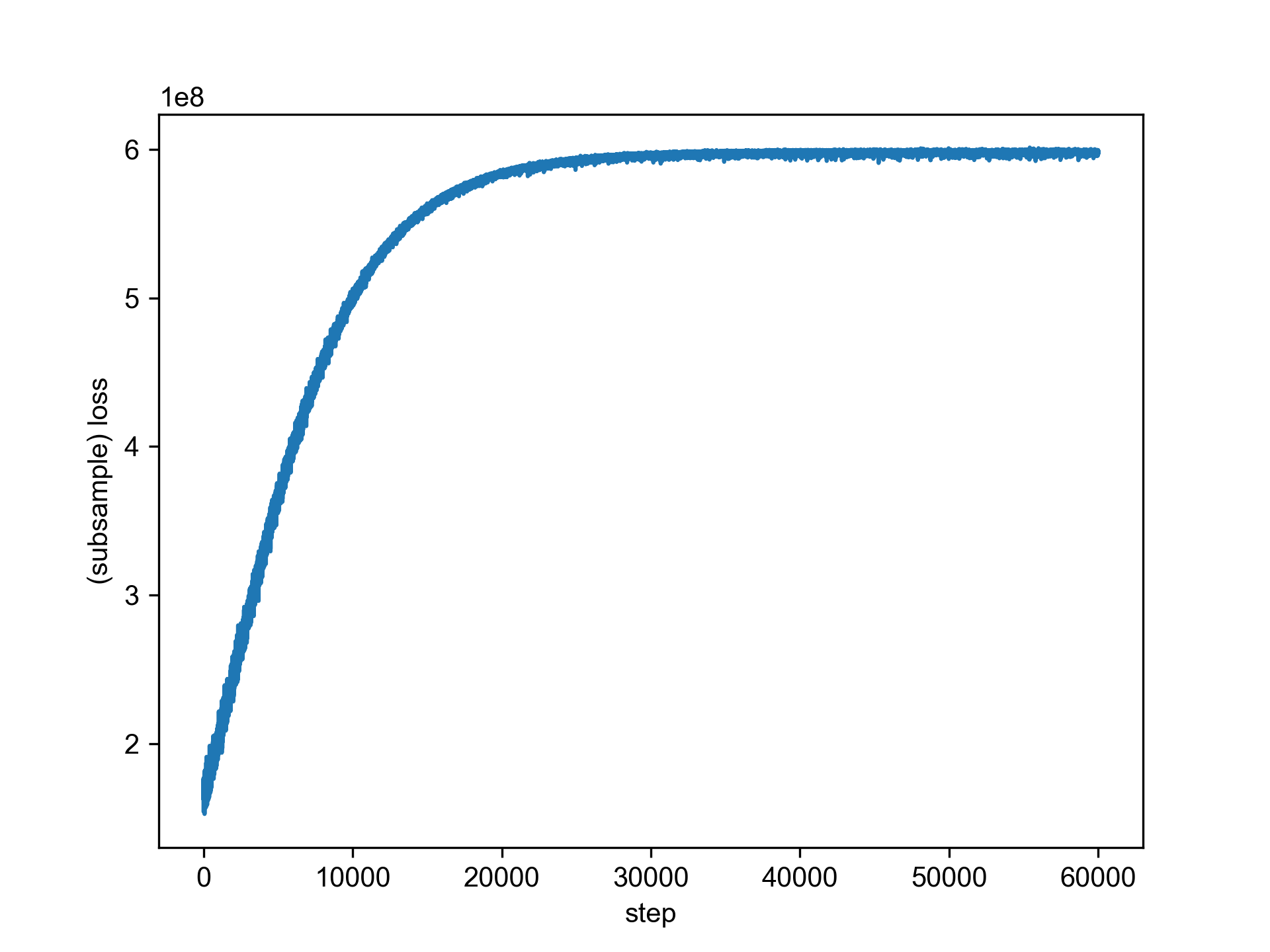}
\end{subfigure}
\begin{subfigure}[t]{0.32\textwidth}
\includegraphics[width=\linewidth]{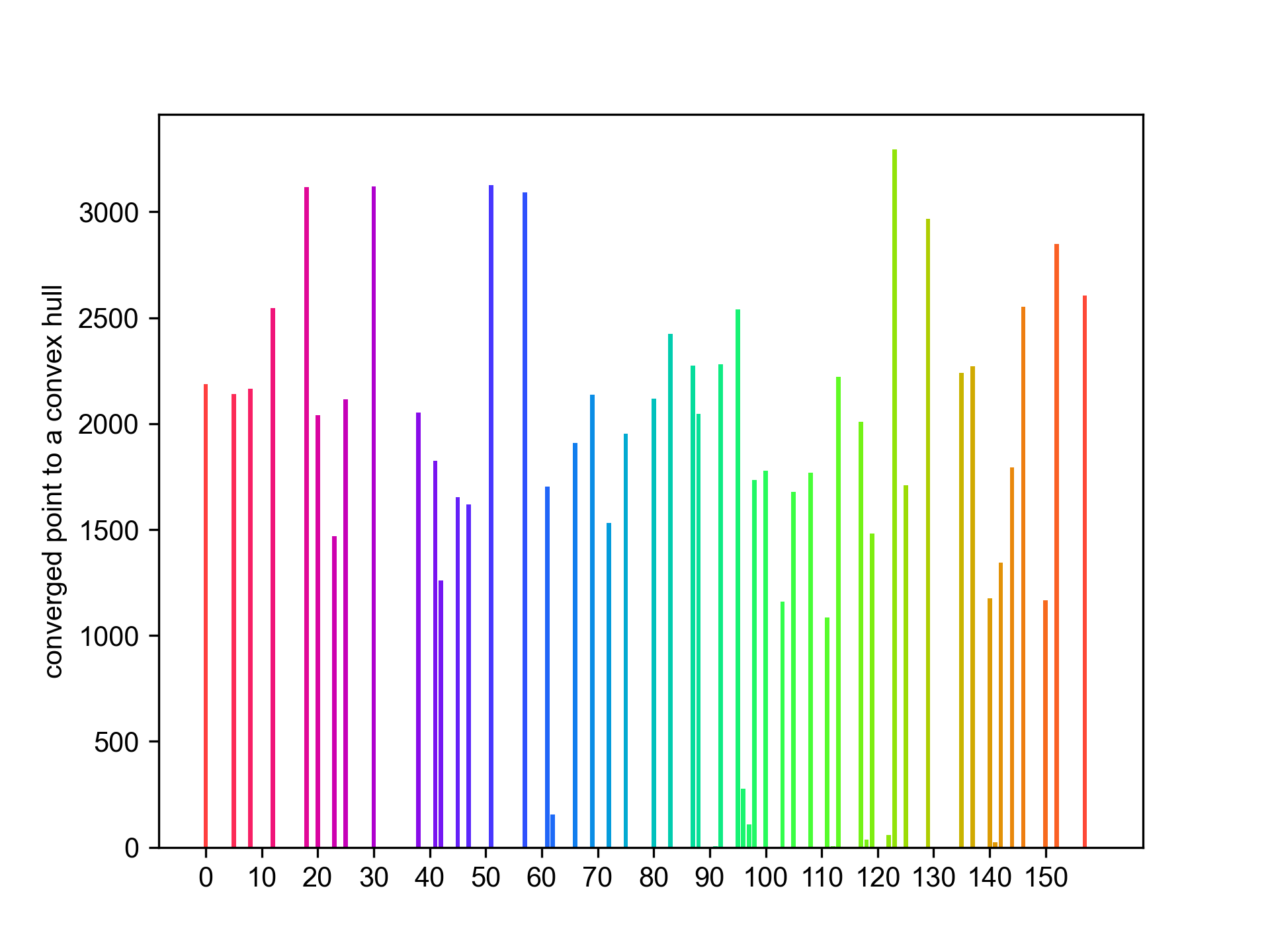}
\end{subfigure}\\\vspace{-15pt}
\caption{Experimental results for the model with $g(z_k-z_l)=|z_k-z_l|^2$ simulated using Alg.~\ref{alg:stoch}, the sub-sample size $S=1000$, $\Delta t=0.02$, and $600$ epochs. The initial agent positions are distributed within a single Gaussian (top), a mixture of $5$ Gaussians (middle), and a ball (bottom). The color coding of the attractor color circles corresponds to the color coding of the attractor density at each convex hull agent position.}
\label{fig:atr}
\end{figure*}

We utilized the following numerical setup to perform all experiments reported in this section by employing Alg.~\ref{alg:stoch}. We fix the pairwise distance function $g(z_k-z_l)=|z_k-z_l|^p$ ($p>0$ is given in the caption), the sub-sample size $S=1000$, time-step $\Delta t=0.02$, and performed $600$ epochs in total. During a single epoch, the communication was computed for all agents. In order to generate the random initial conditions of the agents on the plane that were subject to the dynamical system \eqref{CM}, we used the following protocol. We set the total number of agents spread out according to a mixture of Gaussian distribution with uniform random means and a uniform number of agents in each Gaussian or spread the agents within a ball. 
\begin{algorithm}[h!]
   \caption{Stochastic Simulator}
   \label{alg:stoch}
  
\begin{algorithmic}[1]
   \State\textbf{Input:} total nr of agents $N$; nr of Gaussians; sub-sample size $S$; time step $\Delta t$; number of epochs; distance metric $g$;
   \State\textbf{Output:} agent positions; loss values; the attractor density; 
   \For{number of epochs}
   \State compute ${convex\_hull}\left(\{z_k\}\right)$;\Comment{Qhull\cite{quickhull}}
   \State $\mathcal{N} = shuffle(\{1,\dots,N\})$;
   \For{batch $\mathcal{B}$ of $S$ consecutive indices from $\mathcal{N}$}
   \State compute the communication indices
   \State $l(k)$ for $k\in\mathcal{B}$, and $l\in convex\_hull$ \eqref{def-lk};   
   \EndFor
   \State Update $z_k(t+\Delta t)$ for all $k$ using the dynamics \eqref{CM};
   \EndFor
\end{algorithmic}
\end{algorithm}
\begin{remark}
Due to Thm.~\ref{thm:convex} in Alg.~\ref{alg:stoch}, we can consider only the agents on the convex hull. Thanks to implementing the fast Qhull algorithm \cite{quickhull}, this allows for simulating systems consisting of millions of agents on a laptop (refer to actual wall-times per simulation in Tab.~\ref{tab:compute}. 
\end{remark}
In the first computational experiment, we simulated the agents' dynamics \eqref{SCM} of a large set of agents ($100$k) using a fixed dynamics polarization function $L$ with $g(z_k-z_l)=|z_k-z_l|^2$. We sample the initial $100$k agent positions from three distributions: a single Gaussian, a mixture of Gaussians, and a uniform ball.
The consistency of the attractor obtained by following the stochastic dynamical model described in Sec.~\ref{sec:stochas} is validated by the polarization function value plots demonstrating an apparent stabilization (mid column of Fig.~\ref{fig:atr}). It is clear that the resulting attractors (marked with color dots on the left column of Fig.~\ref{fig:atr})  are contained within the initial convex hull, and the density of each concentration point is presented on the histogram (right column of Fig.~\ref{fig:atr}). After reaching convergence, observe a nonuniform density of agents located at each concentration point. We use the stochastic simulator described in Alg.~\ref{alg:stoch} for simulating the model.

\begin{figure*}[h!]
\captionsetup[subfigure]{justification=centering}
\begin{subfigure}[t]{0.245\linewidth}
\includegraphics[width=\linewidth]{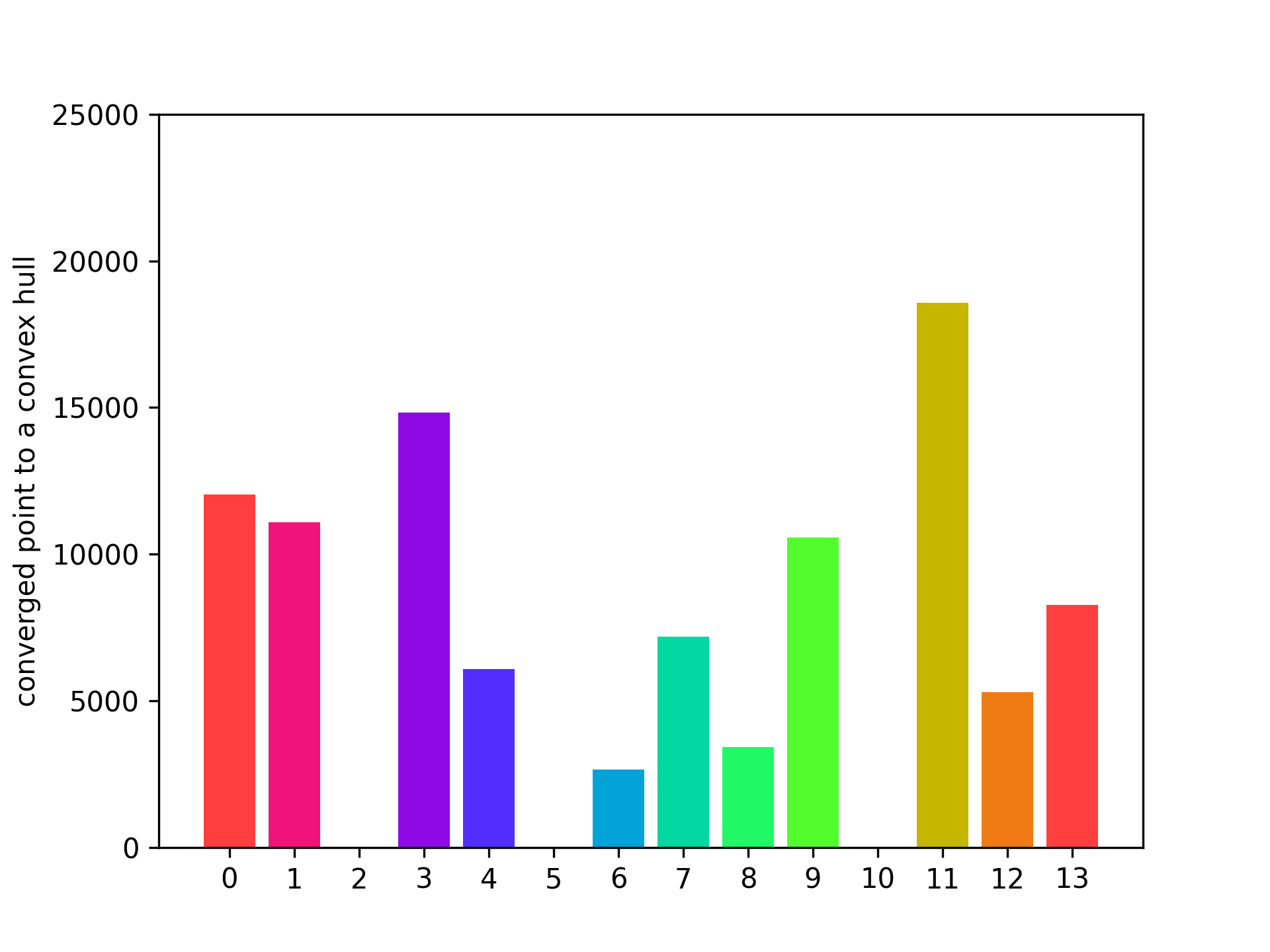}
\caption{$p=1$}
\label{fig:h1}
\end{subfigure}
\begin{subfigure}[t]{0.245\linewidth}
\includegraphics[width=\linewidth]{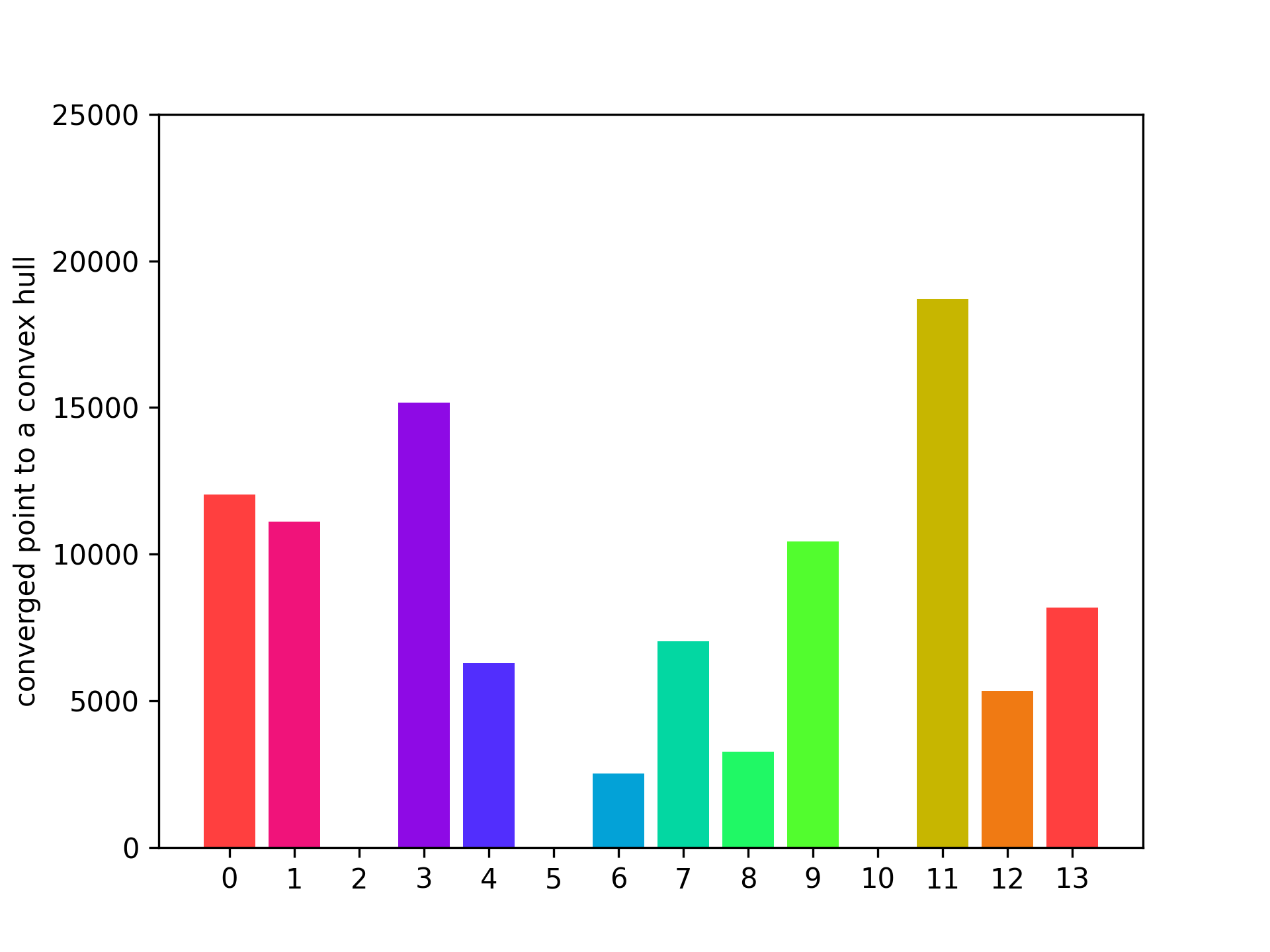}
\caption{$p=2$}
\label{fig:h2}
\end{subfigure}
\begin{subfigure}[t]{0.245\linewidth}
\includegraphics[width=\linewidth]{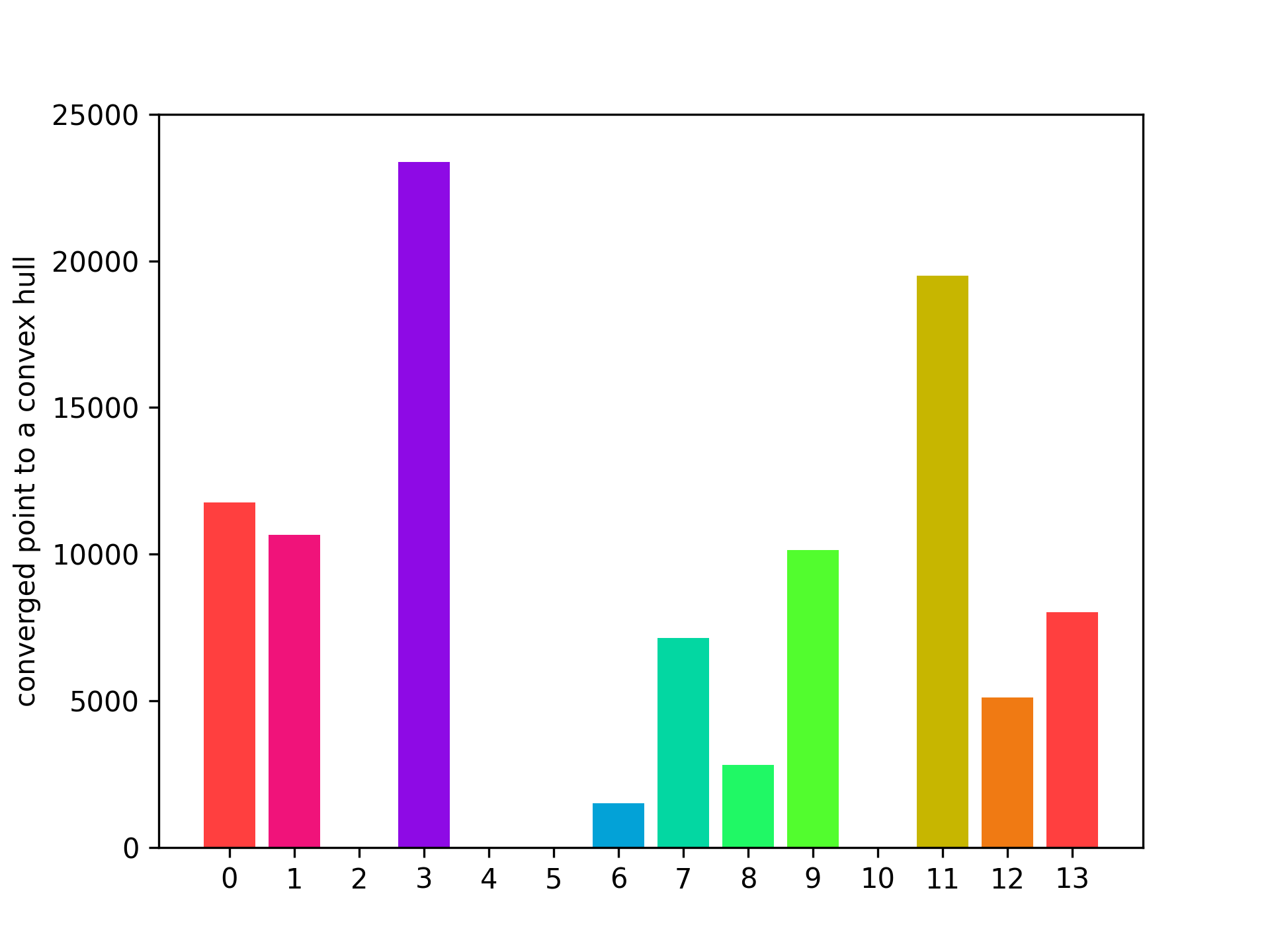}
\caption{$p=4$}
\label{fig:h3}
\end{subfigure}
\begin{subfigure}[t]{0.245\linewidth}
\includegraphics[width=\linewidth]{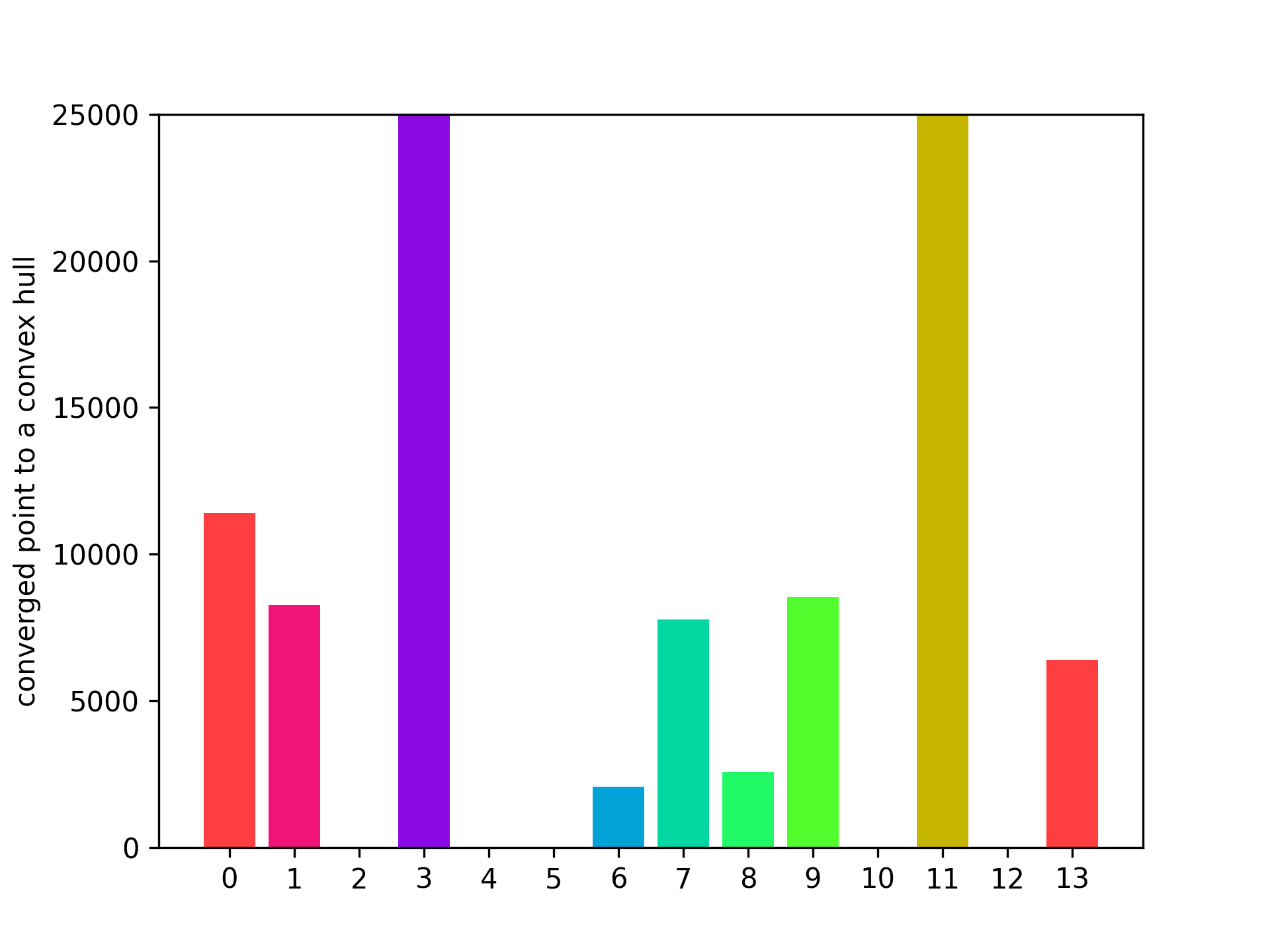}
\caption{$p=10$}
\label{fig:h4}
\end{subfigure}
\caption{Attractor densities of the dynamics defined by using different distance metrics in the polarization function, i.e. $g(z_k-z_l) = |z_k - z_l|^p$.}
\label{fig:pcompare}
\end{figure*}

In the second experiment, we look at the dependence of the dynamical limit on the polarization function; we present the resulting attractors for different forms of $L$ with $g(z_k-z_l) = |z_k-z_l|^p$, setting $p=1, 2, 4, 10$. Initial data is sampled from a single Gaussian. It is observed that the resulting attractor gets concentrated in fewer points the larger the $p$ value is. Results are presented in Fig.~\ref{fig:pcompare}.
\begin{table}[h!]
\caption{Accuracy of the ground truth attractor reconstruction (obtained when $S=N$) in terms of MSE obtained using different sub-sample size $S$ and time-step $\Delta t$ in Alg.~\ref{alg:stoch}. Intial data was sampled from a ball of radius $10$.}
\label{tab:accuracy}
\begin{center}
\begin{sc}
\small
\renewcommand{\arraystretch}{0.75}
\begin{tabular}{cc|cccccc}
\toprule
$N$ & $\Delta t$ & \multicolumn{6}{c}{$S$}\\
 &  & 50 & 100 & 250 & 500 & 1000 & 2500\\\midrule
$5$k & $0.02$ & $17.78$ & $7.11$ & $1.43$ & $0.64$ & $0.18$ & $0.03$ \\
$5$k & $0.01$ & $9.44$ & $4.31$ & $0.96$ & $0.31$ & $0.08$ & $0.02$\\\midrule
$10$k & $0.02$ & $45.82$ & $12.08$ & $2.12$ & $0.68$ & $0.28$ & $0.06$ \\
$10$k & $0.01$ & $13.04$ & $5.63$ & $1.68$ & $0.66$ & $0.24$ & $0.06$ \\\midrule
$20$k & $0.02$ & $60.23$ & $21.88$ & $3.95$ & $1.60$ & $0.47$ & $0.10$ \\
$20$k & $0.01$ & $15.15$ & $6.77$ & $2.47$ & $0.98$ & $0.39$ & $0.09$ \\
\bottomrule
\end{tabular}
\end{sc}
\end{center}
\end{table}

We study the accuracy of the final attractor reconstruction to validate our choice in simulations of the sub-sample $S=1000$ in Alg.~\ref{alg:stoch}. Results are reported in Tab.~\ref{tab:accuracy}. We use the set-up in which initial data is sampled from a ball of radius $10$. Then, the ground truth attractor is obtained by setting $S=N$, and an approximated attractor is obtained by applying Alg.~\ref{alg:stoch} with different subsample $S$. Mean-squared error (MSE) is computed and reported in the corresponding cell in Tab.~\ref{tab:accuracy}. We find, in particular, that $S=1000$ is a choice providing a good compromise between accuracy and runtime. Achieving more accurate simulations with smaller $S$ would require decreasing time-step $\Delta t$ significantly.

In the following experiment, we validate that the sub-sample size $S=1000$ is efficient. 
The computation time increases rapidly with increasing $S$; refer to Tab.~\ref{tab:compute}. We also demonstrate that using such an approach, we are to handle a vast number of agents, and it is computationally cheaper to simulate $N,S=(1000k, 500)$ than $(100k, 5k)$.

\begin{table}[h!]
\caption{Total wall time required to perform $150$ epochs (passes over the full set of agents) of Alg.~\ref{alg:stoch} for various choices of the sub-sample size $S$ and the total number of agents $N$.}
\label{tab:compute}
\begin{center}
\begin{sc}
\small
\renewcommand{\arraystretch}{0.75}
\begin{tabular}{ccc}
\toprule
$S$ & $N$ & wall-time in secs.\\
\midrule
$500$ & $100$k & $498$ \\
$1000$ & $100$k & $712$ \\
$2000$ & $100$k & $1382$ \\
$5000$ & $100$k & $5720$ \\
\midrule
$500$ & $1000$k & $4260$\\
\bottomrule
\end{tabular}
\end{sc}
\end{center}
\end{table}
\section{Predicting Attractor by  Encoder-Decoders}
\label{sec:attractor}
Next, we evaluate the modern approach for predicting the agents' configuration limit by leveraging the predictive capabilities of neural networks. If successful, such an approach can significantly accelerate the attractor computation. We trained and tested the popular deep neural networks in predicting the agent's attractor structure, given solely the initial data. We apply a deep encoder-decoder approach. Given training pairs $\mathcal{D}_{train}=(X_{train},Y_{train})$ ($D_{train}$ pairs in total) and test pairs $\mathcal{D}_{test}=(X_{test},Y_{test})$ ($D_{test}$ pairs in total), a deep neural network $f_\theta(X) = \hat{Y}$ parametrized with a vector of parameters $\theta\in\R^{D_f}$ (its dimension depends on the particular architecture) is trained on $\mathcal{D}_{train}$ by minimizing (stochastic optimization) the MSE loss of the ground truth attractor reconstruction 
$$\mathcal{L}(\theta) = \frac{1}{D_{train}N}\sum_{i=1}^{D_{train}}| f_\theta(X_i) - Y_i|^2.
$$

Observe that the prediction model should work under imposed invariance for the agent permutations. We evaluated two approaches for attaining the invariance for agent permutations. First, we apply a widely used transformer encoder involving self-attention \cite{transformer}. 
Second, we compute a histogram of the agent numbers on a $2$d uniform grid. Hence, the agent positions are encoded as $2$d image, each pixel representing the number of agents. In the actual experiments, we used a $64\times 64$ grid. See  Fig.~\ref{fig:ac}. We train two architectures on the derived dataset. The first uses the self-attention encoder and is applied to the raw agent coordinates and outputs each agent's target positions. The second architecture is inspired by the UNet \cite{unet} encoder-decoder and applied to the 2D agent histograms. We denote the architectures as \verb|attention| and \verb|UNet| respectively.
\subsection{Training \& Test Set Documentation}
For training the models, we crafted a custom dataset of training pairs $(X_{train},Y_{train})$ and test pairs $(X_{test},Y_{test})$ available upon request. 
For each $k$ $(X_{*,k},Y_{*,k})$, $*=train/test$, dimension of $X_{*,k}$ and $Y_{*,k}$ is equal to $200$ (flattened $100\times2$). We applied the following procedure independently, generating each pair.
Draw random (uniform) positions of $N=100$ agents within a ball. Set input $X_{*,k}$ to the flattened vector of agents' positions;
Obtain the ground truth attractor by simulating the agents for $200$ epochs using Alg.~\ref{alg:stoch} (deterministic variant in which $S=N$) with fixed $p=2$, $dt=0.05$. Save the ground truth attractor as the label $Y_{*,k}$. The generated training set contains $D_{train} = 100k$ train pairs and $D_{test}=2.5k$ test pairs.

\subsection{Results Analysis}
\begin{figure*}[h!]
\captionsetup[subfigure]{justification=centering}
\begin{subfigure}[t]{0.245\linewidth}
\includegraphics[width=\linewidth]{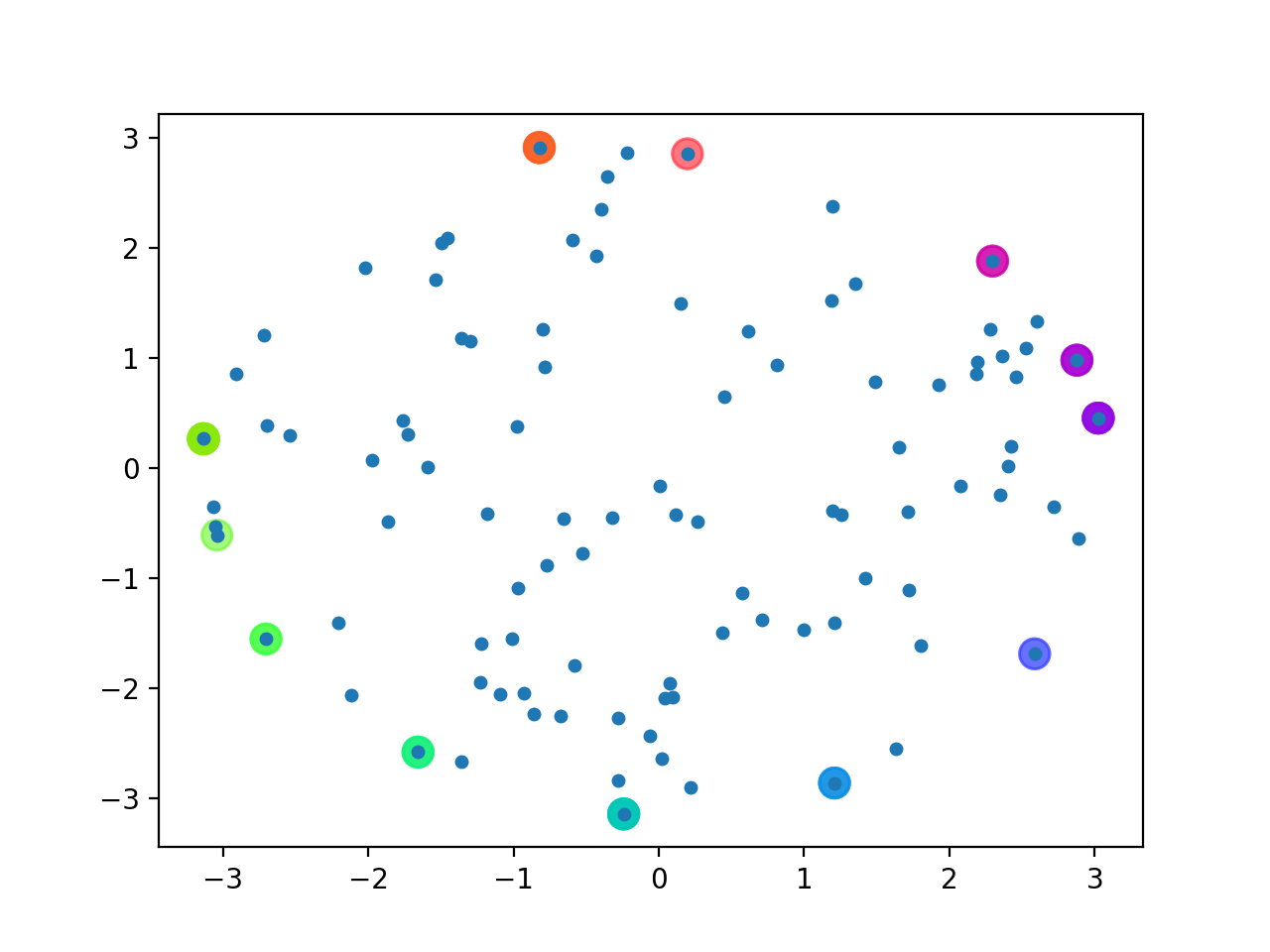}
\caption{ground truth attractor and initial position (agents coordinates)}
\label{fig:agt}
\end{subfigure}
\begin{subfigure}[t]{0.245\linewidth}
\includegraphics[width=\linewidth]{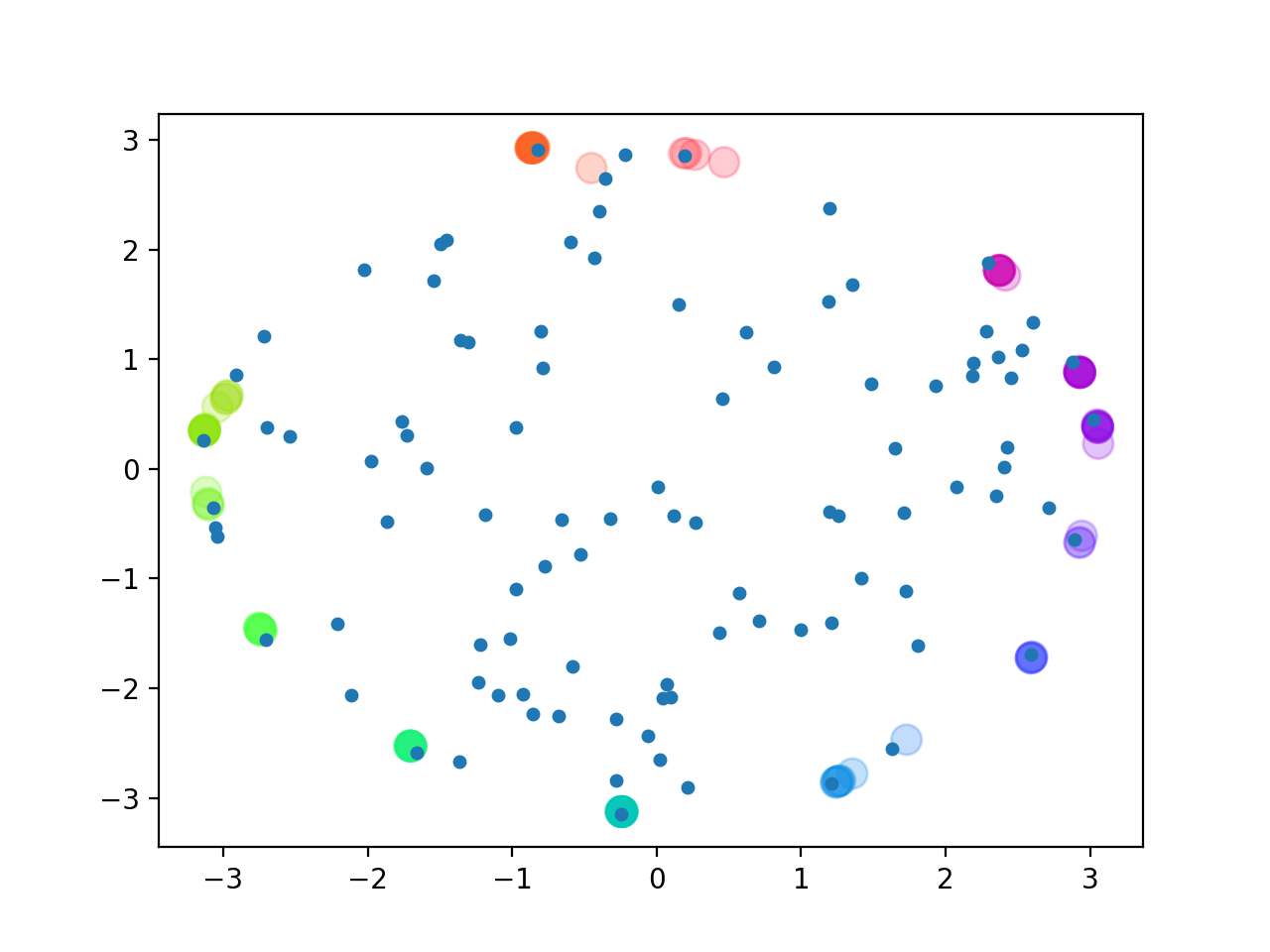}
\caption{\texttt{attention} attractor corresponding prediction (agents coordinates)}
\label{fig:apred}
\end{subfigure}
\begin{subfigure}[t]{0.245\linewidth}
\includegraphics[width=\linewidth]{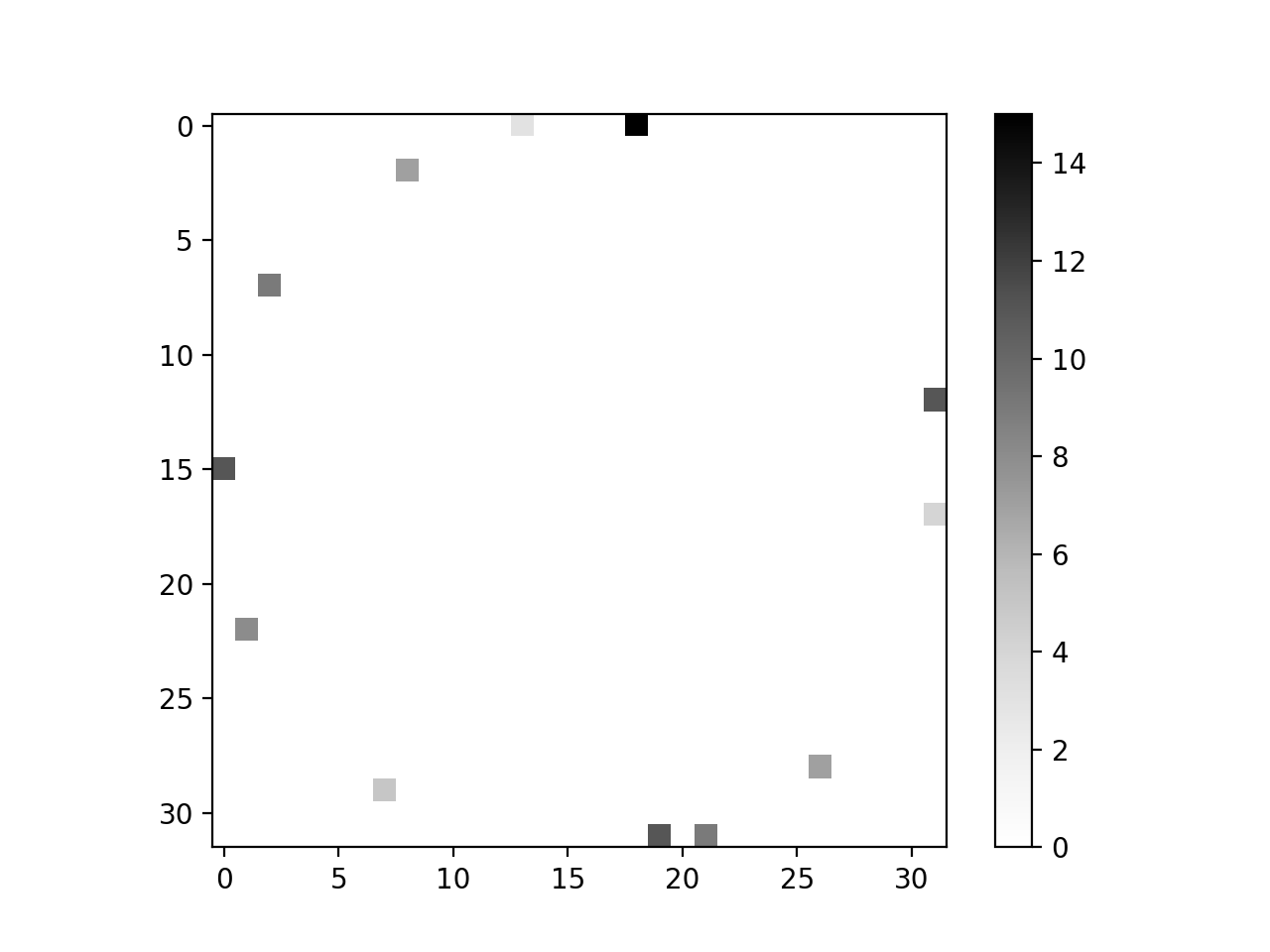}
\caption{ground truth attractor (agents histogram)}
\label{fig:groundtruth}
\end{subfigure}
\begin{subfigure}[t]{0.245\linewidth}
\includegraphics[width=\linewidth]{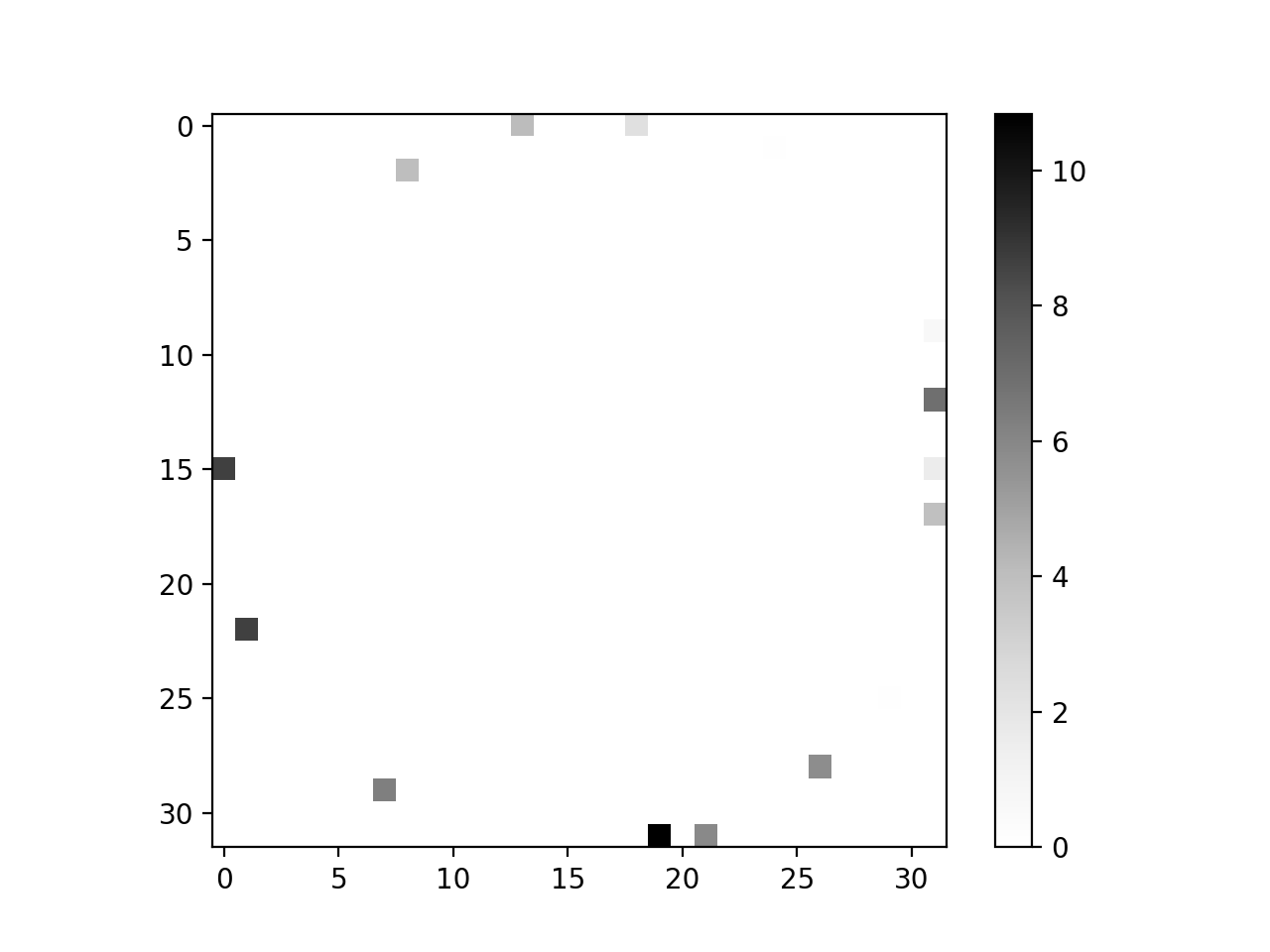}
\caption{\texttt{UNet} corresponding prediction (agents histogram)}
\label{fig:ourpred}
\end{subfigure}
\caption{Ground truth attractor predictors using the two tested models for a sample test initial condition (same for both models). The two plots to the left compare the ground truth with the \texttt{attention} model prediction, achieving MSE error  $0.007$, which is better than the average over all samples $(0.02)$. The two plots to the right present the agents' positions discretized to $32\times 32$ histogram and compared the ground truth with the \texttt{UNet} model prediction, achieving absolute error $41$ (histogram disagreement), which is better than the average over all samples $(80)$. }
\label{fig:ac}
\end{figure*}
We trained the models until each model's test error curve stabilized (models had different losses for training). We did not observe overfitting as the test error along the training kept decreasing, confirming that there is enough 
training 
samples. There is room for improvement in fine-tuning the tested models or other architectural choices to achieve better results, but this is beyond the scope of the present paper; our motivation was initially to validate the utility of neural net models in the given context for further studies.

To interpret the trained models, we cherry-picked one sample illustrating the differences between the two tested models, $x\in X_{test}, y\in Y_{test}$ and the prediction $f_\theta(x)$ for both of the tested autoencoders is in Fig.~\ref{fig:ac}. Note that both models worked on this example better than the average computed over all test samples; the exact loss values are given in the caption. \verb|attention| model correctly captures all the concentration points and their density; however, singular agents are mapped to a wrong position on the outlier, however close to the ground truth point. Predicted attractor by \verb|UNet| model captures all but one ground truth concentration point and adds a few spurious singular points to the attractor that are not present in the ground truth. These are single points and can be easily filtered out. The exact number of agents in \verb|our| prediction vs. the ground truth is slightly off. 
\subsection{Hyperparameters and Model Details}
The models are implemented in the PyTorch framework. Below, we report the technicalities of the models.

\emph{\texttt{attention} model architecture.} The first model is inspired by transformer encoders applied for computing token sequence embeddings. The encoder takes as input the agents' coordinates encoded by a linear layer to the latent vector of dim. $128$. Encoder comprises $8$ layers of stacked self-attention and a feedforward. Self-attention uses $8$ heads, and feedforward is $128$ dimensional. Activation is the ELU. The decoder is a linear layer decoding the latent dimension vector $128$ to the agent's coordinates. We use the Adam optimizer for training, batch size $128$, learning rate $1e-03$, and total $300$ training epochs. The model was trained until the test MSE loss stabilized, the final MSE train error $=0.0185$, and the final MSE test error $=0.016$.

\emph{\texttt{UNet} model architecture}. 
The second model, inspired by the UNet, takes the flattened agent histograms as input and is characterized by the encoder network with three fully connected layers with ReLU activation, $1$d batch normalization, and hidden sizes $=4000$.  We do not apply convolutional layers as we discovered that vanilla UNet performs poorly on the test set and is prone to overfitting.
The final layer outputs a latent representation of size $100$ (equal to $N$). The decoder network is the mirrored encoder. During the encoding process, the autoencoder implements a skip connections mechanism.
The skip connections connect the stored activations from the encoder layers with the corresponding decoder layer. We use the Adam optimizer for training, batch size $512$, learning rate $1e-05$, and total $500$ training epochs. The model was trained until the test mean absolute error (MAE) loss stabilized, the final test MAE $19$, and the train MAE $80$. 
%
%
%

%
%
\section{Conclusions and Future Work}
The present work introduces a simple agent model with strong nonlocal and unstable dynamics, which can be interpreted as a decision-making process for opinions in a 2D scenario. This feature yields fascinating dynamics despite its mathematical ill-posedness at $\Delta t \to 0$. Analyzing  polarization function's dynamics allows us to significantly reduce the computational cost of each time step, limiting the consideration set to the convex hull's vertices. Consequently, we develop a highly efficient solver working for millions of agents while significantly reducing the set into a smaller one, as depicted in Fig. \ref{fig:atr}. 

Future research may explore the impact of choosing a more complex structure for $L(\cdot)$ or function $g(\cdot)$ on the dynamics, which may lead to different outcomes and possibly find a place in data clustering. Additionally, the model's mathematical properties, which hold for any dimension, including pure graph cases, could be leveraged in data reduction problems even for the large dimension case. Finally, further work is needed to develop better ML techniques for non-smooth, non-local, and non-stable dynamical systems.

Our experiment demonstrated that machine learning for recovering agents' dynamics is data-hungry. Training attractor predicting models to achieve good accuracy for many agents is a challenging task requiring a large amount of data. We observed that some of the initial conditions are equally challenging for both of the models. Achieving qualitatively close predictions with only $100$ agents required applying sophisticated deep neural networks, a lengthy training process, and an extensive training dataset of $100$k samples. We leave it as an open question: What dataset sizes are required to train a predictive model for a large set of agents?
%

%

\bibliographystyle{plain}        
\bibliography{autosam}           

\end{document}